\documentclass[reprint,onecolumn,aps,showpacs,nofootinbib,notitlepage]{revtex4-2}
\pdfoutput=1
\usepackage[normalem]{ulem}
\usepackage{graphicx}
\usepackage{psfrag}
\usepackage{caption}
\usepackage{subcaption}
\usepackage{float}
\usepackage{amsmath}
\usepackage{amssymb}
\usepackage{amsthm}
\usepackage{mathtools}
\usepackage{dsfont}
\usepackage{xcolor}
\usepackage{comment}
\usepackage{caption}
\usepackage{bigints}
\usepackage{geometry}
\usepackage{scalerel}[2016/12/29]

\usepackage{lineno}
\usepackage{ulem}

\newcommand{\be}{\begin{equation}}
\newcommand{\ee}{\end{equation}}

\def\bc{\begin{center}}
\def\ec{\end{center}}
\def\bea{\begin{eqnarray}}
\def\eea{\end{eqnarray}}

\newcommand{\deletetext}[1]{\iffalse{{\color{red}{#1}}}\fi}
\newcommand{\newtext}[1]{{\color{black}{#1}}}

\begin{document} 

\title{Accelerated \deletetext{escape}\newtext{first-passage} dynamics in \newtext{a} non-Markovian \deletetext{stochastic}feedback \newtext{Ornstein--Uhlenbeck process}}

\author{Francesco Coghi}
\email{francesco.coghi@nottingham.ac.uk}
\affiliation{School of Physics and Astronomy, University of Nottingham, Nottingham NG7 2RD, United Kingdom}
\affiliation{Nordita, KTH Royal Institute of Technology and Stockholm University, SE-106 91 Stockholm, Sweden}

\author{Romain Duvezin}
\affiliation{Universit\'{e} Paris Saclay, ENS Paris Saclay, 91190 Gif-sur-Yvette, France}
\affiliation{Nordita, KTH Royal Institute of Technology and Stockholm University, SE-106 91 Stockholm, Sweden}

\author{John S.\ Wettlaufer}
\email{john.wettlaufer@yale.edu/jw@fysik.su.se}
\affiliation{Yale University, New Haven, 06520, CT, USA}
\affiliation{Nordita, KTH Royal Institute of Technology and Stockholm University, SE-106 91, Stockholm, Sweden}

\date{\today}


\begin{abstract}

We study the \deletetext{escape}\newtext{first-passage} dynamics of a non-Markovian stochastic process with time-averaged feedback, which we model as a one-dimensional Ornstein--Uhlenbeck process wherein the \newtext{particle} drift is modified by the empirical mean of its trajectory. This process maps onto a class of self-interacting diffusions. Using weak-noise large deviation theory, we \newtext{calculate the leading order asymptotics of the time-dependent distribution of the particle position}, derive the most probable \deletetext{escape}paths \newtext{that reach the specified position at a given time} and quantify their likelihood via the action functional. We compute the feedback-modified Kramers rate and its inverse, which approximates the mean \deletetext{escape}\newtext{first-passage} time, and show that the feedback accelerates \deletetext{escape}\newtext{dynamics} by storing finite-time fluctuations, thereby lowering the effective energy barrier and shifting the optimal \deletetext{escape}\newtext{first-passage} time from infinite to finite. Although we identify alternative mechanisms, such as slingshot and ballistic \deletetext{escape} trajectories, we find that they remain sub-optimal and hence do not accelerate \deletetext{escape}\newtext{the dynamics}. These results show how memory feedback reshapes rare event statistics, thereby offering a mechanism to potentially control \deletetext{escape}\newtext{first-passage} dynamics.

\end{abstract}

\maketitle


\section{Introduction}
\label{sec:Intro}

Many real-world systems exhibit memory-feedback dynamics~\cite{Beran2013, Scalliet2019, Zhang2019, Song2024}, where the evolution of a process depends on its past trajectory. Unlike standard Markovian systems, these processes are influenced by accumulated memory. Such feedback mechanisms introduce rich dynamical features, altering long-time behavior and fluctuation properties in ways not captured by memoryless models~\cite{Schutz2004, Rebenshtok2007, Maes2009, Harris2009, VanMieghem2013, Harris2015, Jack2020, Moran2020, Hartich2021}. In particular, memory-feedback mechanisms are relevant for designing control strategies to achieve specific tasks~\cite{Sutton1998, Bechhoefer2005, Puterman2008, Khadem2019, Bechhoefer2021} with specific thermodynamic properties~\cite{Horowitz2010, Sagawa2012, Jun2014, Loos2019, Rico-Pasto2021, Debiossac2022, Kopp2024}.

Self-interacting systems~\cite{Toth2001, Benaim2002, Pemantle2007, DelMoral2007, Kurtzmann2010, Brmont2024, Coghi2025} constitute a notable class of memory-feedback processes in which past states, encoded in the empirical occupation measure, or in a time-integrated functional of the process, influence future evolution. Examples include models in statistical physics relevant to biological~\cite{Budrene1995, Tsori2004, Reid2012, Zhao2013, Kranz2016} and colloidal systems~\cite{Nakayama2023, Kumar2024, Chen2025}, where interactions with the occupation measure affect the dynamics.


Although the stationary properties of self-interacting processes have been studied extensively, the role of feedback in shaping rare event statistics, and in particular \deletetext{escape}\newtext{first-passage} dynamics, remains poorly understood. However, understanding the finite-time stochastic dynamics of feedback-driven systems has many important implications in, among other problems, reinforcement learning algorithms~\cite{Sutton1998}.


Only recently have large-deviation results for self-interacting systems begun to emerge, both in the long-time regime for discrete Markov chains~\cite{Budhiraja2022, Budhiraja2023} and for diffusions~\cite{Coghi2024}. For example, \deletetext{escape}\newtext{first-passage} dynamics has been addressed in self-interacting discrete-time random walks~\cite{Barbier-Chebbah2022}, where universal scaling laws for \deletetext{escape}\newtext{first-passage} time distributions were identified, and in diffusive systems~\cite{Aleksian2024}, where by analyzing the asymptotic behavior of the \deletetext{escape}\newtext{first-passage} time in the weak-noise limit a Kramers-type law was derived.


Here we analyze a representative case: a one-dimensional Ornstein--Uhlenbeck process where the drift term is influenced by its own time-averaged position. This feedback mechanism breaks Markovianity and defines a self-interacting diffusion process, fundamentally altering the \deletetext{escape}\newtext{first-passage} dynamics of the system.


By using weak-noise large deviation theory, we show that self-interaction can lead to accelerated \deletetext{escape}\newtext{first-passage dynamics}, and thus fluctuation-driven mechanisms induced by memory. In particular, we identify and characterize different \deletetext{escape}\newtext{first-passage} scenarios and quantify their likelihood. Our large deviation approach allows us to derive a Kramers-type law for the mean \deletetext{escape}\newtext{first-passage} time in the feedback Ornstein--Uhlenbeck process, which complements and extends the results in~\citep{Aleksian2024, Barbier-Chebbah2022}. \newtext{We show that for a positive feedback parameter $\beta>0$ (see Eq.\ \eqref{eq:Process}), the leading-order noise asymptotics of the first-passage time distribution develops a new global minimum in time. As a result, the effective energy barrier is reduced, causing exponential speed-up and an exponentially smaller mean first-passage time.} Furthermore, our findings generalize recent results on \deletetext{escape}\newtext{first-passage} dynamics in other classes of memory-driven systems, such as generalized Langevin equations~\cite{Kappler2018, Barbier-Chebbah2024}.


The paper is organized as follows.  In Section~\ref{sec:Model}, we introduce the Ornstein--Uhlenbeck process with empirical average position feedback, establish its mapping to self-interacting diffusion, and discuss its deterministic and long-time dynamics.
In Section~\ref{sec:Weak}, we present the weak-noise large deviation framework used to derive our results. We apply it to both the standard and feedback Ornstein--Uhlenbeck processes, and derive the fundamental equation~\eqref{eq:xtFinEq} governing finite-time dynamics in the weak-noise regime, which we have not found in the literature. In Section~\ref{sec:Results}, we detail the consequences of solving~\eqref{eq:xtFinEq}. We describe the finite-time mechanisms driving \deletetext{escape}\newtext{first-passage dynamics} in the feedback process, compare them to the standard Ornstein--Uhlenbeck case, and analyze their relative and absolute likelihoods. We characterize the quasi-potential (i.e., the large deviation rate function associated with the Arrhenius law) and the optimal \deletetext{escape}\newtext{first-passage} times.  
Finally, in Section~\ref{sec:Conclusion} we summarize our findings, highlight open questions, and discuss potential applications, particularly in the context of reinforcement learning.

\section{Stochastic Feedback Ornstein--Uhlenbeck Processes and Deterministic Dynamics}
\label{sec:Model}

\subsection{Ornstein--Uhlenbeck Process with Empirical Average Position Feedback}
\label{subsec:OUfeedback}

We consider a one-dimensional process $(X_t)_{t \geq 0}$ governed by the following stochastic differential equation (SDE):
\begin{equation}
\label{eq:Process}
dX_t = - X_t \, dt + \frac{\beta}{t} \left( \int_0^t X_s \, ds \right) dt + \sqrt{\epsilon} \, dW_t \, ,
\end{equation}
with initial condition $X_0 = 0$, noise intensity $\epsilon > 0$, and feedback parameter $\beta \in (-\infty, 1)$, the range of which we will justify in detail below. 
Here, $W_t$ denotes a standard Wiener process, or Brownian motion, in $\mathbb{R}$, characterized by independent Gaussian increments: $\mathbb{E}[dW_t] = 0$ and $\mathbb{E}[dW_s dW_t] = \delta(t-s)$.

In the absence of the additional feedback term involving the time-averaged position $\frac{1}{t} \int_0^t X_s \, ds$, Eq. \eqref{eq:Process} is the classical Ornstein--Uhlenbeck (OU) process, describing a Brownian agent moving in a parabolic potential well centered on the origin. As we will show, the inclusion of the empirical time-averaged position introduces a non-Markovian feedback that can give rise to unique dynamical behavior, effectively modifying the potential in which the particle evolves. This results in an \emph{effective potential} whose minimum shifts along $\mathbb{R}$ depending on the cumulative history of the process.  We discuss this in more detail in Section~\ref{subsec:Mapping}.

Our objective is to analyze how this non-Markovian feedback influences the \deletetext{escape}\newtext{first-passage} dynamics of $X_t$ through a threshold $x_f > 0$, with a particular focus on the role of noise and memory. The feedback term can be expressed as
\begin{align}
\label{eq:FeedbackDefinition}
\bar{X}_t &\coloneqq \frac{1}{t} \int_0^t X_s \, ds \\
\label{eq:FeedbackEmpirical}
&= \int_{\mathbb{R}} x \, \rho_t(x) \, dx \,,\qquad \textrm{where}
\end{align}
\begin{equation}
\label{eq:EmpiricalOccupation}
\rho_t(x) = \frac{1}{t} \int_0^t \delta(X_s - x) \, ds
\end{equation}
is the empirical occupation measure of the process. This measure captures the fraction of time the system has spent in the vicinity of each position $x$ up to time $t$.

Although because its evolution depends on its entire history, the process $(X_t)_{t \geq 0}$ is non-Markovian, the joint two-dimensional process $(X_t, \bar{X}_t)_{t \geq 0} \eqqcolon (Z_t)_{t \geq 0}$ is Markovian and satisfies:
\begin{align}
\label{eq:ProcessMarkov}
\begin{cases}
dX_t &= - X_t \, dt + \beta \bar{X}_t \, dt + \sqrt{\epsilon} \, dW_t \\
d\bar{X}_t &= \frac{X_t - \bar{X}_t}{t} \, dt \, .
\end{cases}
\end{align}
Despite the fact that the process $(Z_t)_{t \geq 0}$ is time-inhomogeneous, the Markovian formulation \eqref{eq:ProcessMarkov} allows us to study the \deletetext{escape}\newtext{first-passage} dynamics of the feedback OU process semi-analytically in the weak-noise regime, as shown in Section~\ref{subsec:WeakFeedbackOU}.

\subsection{Mapping to Self-Interacting Diffusion Processes}
\label{subsec:Mapping}

The structure of the feedback embodied in Eqs.~\eqref{eq:FeedbackDefinition}--\eqref{eq:EmpiricalOccupation}, allows the process $(X_t)_{t \geq 0}$ in Eq.~\eqref{eq:Process} to be interpreted as a specific case of a one-dimensional self-interacting diffusion (SID). A general SID in $\mathbb{R}^d$ is typically expressed as
\begin{equation}
\label{eq:SID}
dX_t = - \nabla V(X_t) \, dt - \frac{1}{t} \left( \int_0^t \nabla K(X_s, X_t) \, ds \right) dt + \sqrt{\epsilon} \, dW_t \, ,
\end{equation}
where $V(X_t)$ is a background potential and $K(X_s, X_t)$ is an interaction potential whose gradient is taken with respect to $X_t$. The interaction term describes how the current position is influenced by the history of past positions\newtext{, functionally through the empirical occupation measure defined in Eq.\ \eqref{eq:EmpiricalOccupation}}.

\newtext{The long-time behavior of self-interacting processes, including random walks and diffusions, has been an active area of research in probability theory for several decades. Self-interacting random walks were first introduced in the discrete-time setting by Norris et al.~\cite{Norris1987}, and continuous-time counterparts appeared soon after as models for polymer growth~\cite{Durrett1992}. The \emph{normalized} SID considered here (see Eq.~\eqref{eq:SID}) was introduced by Benaïm et al.~\cite{Benaim2002} as an example of a continuous-time path-interaction (or reinforcement) process. It extends earlier classes of self-interacting processes based on non-normalized occupation measures~\cite{Cranston1995,Raimond1997} and discrete-time reinforced random walks.  A thorough historical overview is given in \cite{Pemantle2007}.  

The asymptotic properties of SIDs were first studied in~\cite{Benaim2002,Benaim2003,Benaim2005,Benaim2011}, with results established for compact spaces and symmetric interaction potentials. These were later extended to open domains, in particular to $\mathbb{R}^d$ under confining potentials~\cite{Kurtzmann2010,Chambeu2011,Kleptsyn2012}, which is directly relevant for our feedback OU process. Stochastic approximation methods were used in~\cite{Benaim2002,Kurtzmann2010} to describe the asymptotic behavior of the occupation measure $\rho_t$ via a limiting non-autonomous differential equation. A key result of~\cite{Benaim2005} shows that for symmetric potentials $\rho_t$ converges almost surely to a local minimum of a nonlinear free-energy functional, which typically admits multiple critical points, each selected at random with some probability. In the special case of symmetric self-repelling potentials this functional is strictly convex, so $\rho_t$ converges almost surely to a unique global minimum in the long-time limit.  

More recently, refinements of Kramers’ law for SIDs have been obtained~\cite{Aleksian2022,Aleksian2024} (see also~\cite{Kappler2018,Barbier-Chebbah2024} for generalized overdamped Langevin systems), and these results extend to systems with general additive noise~\cite{Aleksian2022,Aleksian2024Thesis,Aleksian2024}. A recent contribution by one of us establishes large-deviation results for SIDs on the ring~\cite{Coghi2024}. We will detail below (Sec.~\ref{subsec:LongTime}) those results that are most relevant for the present work.} 

For the case of Eq.~\eqref{eq:Process}, the potentials in Eq.~\eqref{eq:SID} take the form
\begin{align}
\label{eq:SIDBackgroundFeedback}
V(X_t) &= \left(1 - \beta \right) \frac{X_t^2}{2} \qquad \textrm{and}\\
\label{eq:SIDInteractionFeedback}
K(X_s, X_t) &= K(X_t - X_s) = \beta \frac{(X_t - X_s)^2}{2} \, .
\end{align}
Therefore, a quadratic interaction potential $K$ describes the attraction of the current position $X_t$ to the past positions $X_s$, so that the general behavior of the process is determined by the combination of this interaction and the background potential $V$.

By rewriting Eq.~\eqref{eq:Process}, we can define an effective potential
\begin{equation}
\label{eq:EffectivePotentialFeedbackOU}
\tilde{V}(X_t) = V(X_t) + \frac{1}{t} \int_0^t K(X_s, X_t) \, ds \, ,
\end{equation}
whose shape evolves according to the past trajectory of the process. For example, the minimum of this effective potential may shift to the left or right of the origin depending on whether the integral contribution in Eq.~\eqref{eq:EffectivePotentialFeedbackOU} is negative or positive, respectively.

Eqs.~\eqref{eq:SIDBackgroundFeedback} and \eqref{eq:SIDInteractionFeedback} make clear that for $\beta < 1$ ($\beta > 1$), the background potential remains quadratic and attractive (repulsive) around the origin.  Although the process is still attracted to its past positions through the interaction potential $K$, the repulsive background potential drives \deletetext{escape}\newtext{first-passage dynamics} towards infinity, which justifies the choice $\beta < 1$ described when we introduced Eq.~\eqref{eq:Process}.


More generally, Eq.~\eqref{eq:Process} can be extended to incorporate alternative forms of feedback, such as
\begin{equation}
\label{eq:GeneralFeedback}
\bar{K}_t \coloneqq t^{-\alpha} \int_0^t f(X_s) \, ds \, ,
\end{equation}
with $\alpha > 0$, which ensures a decaying contribution, with slower decay as $\alpha$ decreases, and $f$ is an arbitrary function on $\mathbb{R}$. Although the methods developed here could be generalized to such circumstances, we focus on the simplest and most natural case: feedback via the empirical average position, corresponding to $\alpha = 1$ and $f(X_s) = X_s$. Importantly, this choice enables an exact mapping to SID processes. Alternative values of $\alpha$, or more general time-dependent functions, would typically not allow this mapping and thus preclude comparison with well-established convergence properties of SIDs, which we discuss in Section~\ref{subsec:LongTime}.

\subsection{Exact Results for the Deterministic Dynamics}
\label{subsec:Deterministic}

Before analyzing the stochastic dynamics of the feedback OU process, we examine some essential features of the deterministic dynamics, obtained by setting $\epsilon = 0$ in Eqs.~\eqref{eq:Process} or \eqref{eq:ProcessMarkov}. This analysis provides insight into the behavior of the system and helps clarify the choice of the parameter range for $\beta$.


We set $\epsilon = 0$ in Eq.~\eqref{eq:ProcessMarkov}, which becomes the following first-order, time-inhomogeneous linear system of differential equations; 
\begin{equation}
\label{eq:LinearDeterm}
\dot{Z}_t = A_t(\beta) Z_t \, ,
\end{equation}
where $\dot{(~~)}$ denotes the time derivative, and the time-dependent drift matrix is
\begin{equation}
\label{eq:MatrixDrift}
A_t(\beta) = 
\begin{pmatrix}
-1 & \beta \\
\frac{1}{t} & - \frac{1}{t} 
\end{pmatrix} \, .
\end{equation}

For $\beta \in \mathbb{R} \setminus \{1\}$, the only fixed point of the system is the origin $(0,0)$. However, when $\beta = 1$, all points satisfying $X = \bar{X}$ are fixed points. 

The stability of the origin depends on both $\beta$ and time as follows \newtext{(see Fig.\ \ref{figphase} for comparison)}:
\begin{itemize}
    \item[$\beta < 0$:] The origin is stable, but exhibits a spiral-like behavior in the finite time window $[t_-, t_+]$ defined below.
    \item[ $\beta \in [0,1)$:] The origin is a stable node.
    \item[$\beta > 1$:] The origin becomes a saddle point, and an instability develops along a specific direction. At the bifurcation point $\beta=1$, all points $X=\bar{X}$ are neutrally stable.
\end{itemize}

\newtext{
\begin{figure}[h!]
    \centering
    \includegraphics[width=\textwidth]{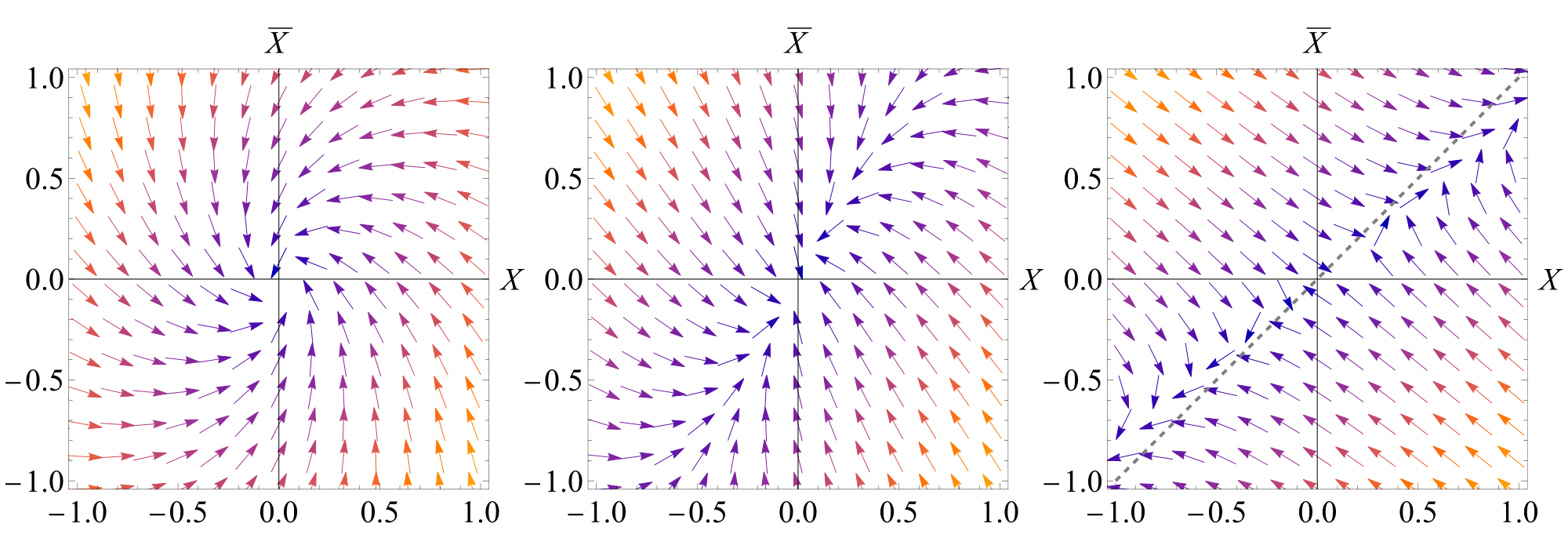}
    \caption{\newtext{Phase portraits of the deterministic dynamical systems in Eq.\ \eqref{eq:LinearDeterm} with \eqref{eq:MatrixDrift} for three illustrative values of $\beta = -0.5$ (left), $\beta = 0.25$ (middle), and $\beta = 1.5$ (right).}}
    \label{figphase}
\end{figure}
}

Due to the linearity of Eq.~\eqref{eq:LinearDeterm}, stability is fully characterized by the eigenvalues of the $2 \times 2$ matrix $A_t(\beta)$, which are always real and given by:
\begin{equation}
\label{eq:EigenvaluesDeterm}
\lambda^{\pm}_t = \frac{-(1 + t) \pm \sqrt{(1 + t)^2 - 4t(1 - \beta)}}{2t} \, .
\end{equation}
For $\beta \in [0,1)$, the square root term in Eq.~\eqref{eq:EigenvaluesDeterm} satisfies:
\begin{equation}
\label{eq:Condition}
\sqrt{(1 + t)^2 - 4t(1 - \beta)} \leq 1 + t \, ,
\end{equation}
implying that both eigenvalues $\lambda_t^\pm$ are negative for all $t > 0$, and the origin is a stable node.

For $\beta < 0$, the discriminant in Eq.~\eqref{eq:EigenvaluesDeterm} can become negative, leading to complex eigenvalues. This occurs when $t \in [t_-, t_+]$, where
\begin{equation}
\label{eq:Conditiont}
t_{\pm} = -(2\beta - 1) \pm 2 \sqrt{\beta(\beta - 1)} \, .
\end{equation}
Since the real parts of both eigenvalues remain negative, the origin remains stable, but the dynamics exhibit a spiral behavior during this time interval.

For $\beta > 1$, the situation is reversed: one eigenvalue $\lambda^-_t < 0$, while the other $\lambda^+_t > 0$. An unstable direction thus emerges, associated with the eigenvector
\begin{equation}
\label{eq:UnstableDirection}
u^+_t = \begin{pmatrix}
\frac{1}{2} \left[ 1 - t + \sqrt{(1 + t)^2 - 4t(1 - \beta)} \right] \\
1
\end{pmatrix} \, .
\end{equation}
Here, the fixed point at the origin becomes a saddle. When $\beta = 1$, the direction $X = \bar{X}$ is neutrally stable with $\lambda^+_t = 0$. Thus, any small perturbation to $(0,0)$ not orthogonal to $u^+_0$ will drive the system away from the fixed point, growing asymptotically along the direction of $u^+_t$, with $X$ and $\bar{X}$ increasing proportionally, as in Eq.~\eqref{eq:UnstableDirection}.

The following features of the deterministic dynamics have the following important implications for the stochastic system. 

\begin{itemize}

\item[(1)] If one conditions on $X_T = x_f > 0$, then when $\beta > 1$ even a small perturbation due to noise will cause the process to escape the origin deterministically towards $x_f$. In such a regime, noise merely initiates the escape, but deterministic dynamics dominate the evolution. \newtext{We will not focus on this regime in our work.  However, it would be interesting to study the fluctuations of the escape time from the origin, similarly to that done in~\cite{Krapivsky2025}.}

\item[(2)] The interesting regime \newtext{for our work} is when $\beta < 1$. For $\beta \in [0,1)$, the feedback in Eq.~\eqref{eq:Process} is positive, so the most likely \deletetext{escape}\newtext{first-passage} to $x_f > 0$ occurs through the accumulation of positive fluctuations. However, for $\beta < 0$, this intuitive picture breaks down: as positive fluctuations accumulate, the feedback becomes increasingly negative, pulling the process back toward the origin and making \deletetext{escape}\newtext{first-passage} through $x_f$ highly unlikely.


Therefore, two alternative \deletetext{escape}\newtext{first-passage} mechanisms for the stochastic dynamics emerge:

\item[(3)] (a) A one-stage mechanism, whereby large fluctuations are sufficient to overcome the negative feedback.  This is an extremely rare event in the weak-noise limit. (b) A two-stage mechanism, whereby the stochastic process first accumulates negative fluctuations, generating a strong positive feedback, which then drives a ``slingshot'' \deletetext{escape}\newtext{first-passage} to the right; $x_f > 0$.


\end{itemize}

These $\beta$-dependent qualitatively distinct \deletetext{escape}\newtext{first-passage} pathways raise a natural question: in a symmetric domain $[-x_f, x_f]$, with $\beta \in [0,1)$, is the agent more likely to escape to the right via the accumulation of positive fluctuations, or to the left by exploiting a slingshot mechanism? We return to these questions in Section~\ref{sec:Results}.


In the next section (\ref{subsec:LongTime}), we characterize the long-time behavior of general SIDs and feedback processes, which will serve as a useful reference for our analysis of finite-time \deletetext{escape}\newtext{first-passage} dynamics.

\subsection{Long-Time Behavior of the Feedback Ornstein--Uhlenbeck Processes}
\label{subsec:LongTime}

Having established in Section~\ref{subsec:Mapping} that the feedback OU process in Eq.~\eqref{eq:Process} belongs to the class of SIDs, we now draw on results from SID theory to understand the important asymptotic behavior. This will prove useful later, particularly in Sections~\ref{subsec:Action} and~\ref{subsec:Quasi-potential}, when interpreting our results.

Assuming stationary behavior at long times (as in~\cite{Benaim2002,Aleksian2022,Coghi2024}), one can formally derive the stationary occupation measure, or invariant density, $\rho_{\text{inv}}$ of the general SID in Eq.~\eqref{eq:SID} from the stationary Fokker--Planck equation (see~\cite{Coghi2024}):
\begin{equation}
\label{eq:StationaryMeasure}
\rho_{\text{inv}}(x) = \frac{e^{- \frac{2}{\epsilon} \left( V(x) + \int_{\mathbb{R}^d} K(y,x) \rho_{\text{inv}}(y) \, dy \right)}}{Z} \, ,
\end{equation}
where $Z$ is a normalization constant. The symmetry of the potential $V$ in Eq.~\eqref{eq:SIDBackgroundFeedback} is inherited by the invariant density $\rho_{\text{inv}}$~\cite{Carrillo2003,Aleksian2022,Aleksian2024}, and in the case of the feedback OU process Eq.~\eqref{eq:Process}, Eq.~\eqref{eq:StationaryMeasure} becomes
\begin{equation}
\label{eq:StationaryMeasureFeedbackOU}
\rho_{\text{inv}}(x) = \sqrt{\frac{1}{\pi \epsilon}} \, e^{- \frac{x^2}{\epsilon}} \, ,
\end{equation}
which coincides with the stationary distribution of the standard OU process. In consequence, the memory feedback term in Eq.~\eqref{eq:FeedbackDefinition} converges in probability to
\begin{equation}
\label{eq:MemoryFeedbackLongTime}
\bar{X}_t \xrightarrow{t \rightarrow \infty} 0 \, .
\end{equation}
Therefore, the classical OU process is recovered as the long-time limit of the feedback OU process.
However, it is important to note that the convergence to this fixed point occurs \emph{very-slowly}; slower than any algebraic rate. This has been shown in~\cite{Kleptsyn2012,Aleksian2022,Aleksian2024}: see Theorems~1.6 and 1.12 in~\cite{Kleptsyn2012}, Theorem~2.7 in~\cite{Aleksian2022}, and Theorem~2.10 and  Propositions~2.11--2.12 in~\cite{Aleksian2024}. These long-time results serve as a valuable benchmark for understanding the finite-time dynamics studied in the remainder of this paper, where the memory feedback has a controlling influence.


We conclude this section by highlighting Theorem 2.13 from Aleksian et al.\ \cite{Aleksian2024}, which is relevant to our study. In the weak-noise limit $\epsilon \rightarrow 0$, they established a Kramers-type law for the \deletetext{escape}\newtext{first-passage time} the process in Eq.~\eqref{eq:SID} exits a domain $\mathcal{D} \subset \mathbb{R}^d$\newtext{, through the boundary $\partial \mathcal{D} \subset \mathbb{R}^d$, defined as} $T_0(x) \coloneqq \inf \{ t \geq 0 : X_t \notin {\cal D} \, | \, X_0 = x \in \mathcal{D} \}$.
\deletetext{probability of the agent  
through an open domain $\cal{D}$ such that the boundary is $\partial {\cal D} \subset \mathbb{R}^d$.} They assumed that the potentials $V$ and $K$ are uniformly convex with at most polynomial growth and $V$ such that $\lim_{|x| \rightarrow \infty} |\nabla V(x)|^2 / V(x) = +\infty$ (this last condition does not hold for the $V$ in Eq.~\eqref{eq:SIDBackgroundFeedback}). \newtext{They established that
\begin{equation}
    \label{eq:KramerSIDTime}
    \lim_{\epsilon \rightarrow0} \frac{\epsilon}{2} \log T_0(x) = \inf_{x' \in \partial \mathcal{D}} H(x') \, ,
\end{equation}}
\deletetext{The \deletetext{escape}\newtext{first-passage} probability distribution can be expressed as:
\begin{equation}
\label{eq:KramerSID}
\mathbb{P}(X_{\tau} \in \partial {\cal D}) \approx \exp \left[ - 2 \epsilon^{-1} \inf _{x \in \partial \mathcal{D}} H(x) \right] \, ,
\end{equation}}
where
\begin{equation}
\label{eq:HDef}
H(x') \coloneqq V(x') + K(x' - m) - V(m) \, ,
\end{equation}
\deletetext{where The first time the process exits the domain $\mathcal{D}$ is $\tau \coloneqq \inf \{ t \geq 0 : X_t \notin \partial {\cal D} \}$,}
and $m$ is the point at which $\rho_{\text{inv}}$ attains its maximum. For the feedback OU process in Eq.~\eqref{eq:Process}, we have $m = 0$ and we typically consider a single \deletetext{escape}\newtext{first-passage} boundary $\partial {\cal D} = \{x_f\}$. Therefore, \deletetext{the rate function becomes}\newtext{we find}:
\begin{equation}
\label{eq:KramerSIDExponent}
H(x_f) = x_f^2 \, .
\end{equation}
We note that this result was derived under the further assumption that the empirical occupation measure stabilizes prior to \deletetext{escape}\newtext{first-passage dynamics}\newtext{, which we believe is the reason the right hand side of \  Eq.\ \eqref{eq:KramerSIDTime} depends neither on the starting point $x$ nor or on $\beta$}. As we show in Section~\ref{subsec:WeakFrame} using a weak-noise large deviation approach, our analysis does not require this assumption, which leads to a strong dependence of the \deletetext{escape}\newtext{first-passage} dynamics on the value of $\beta \in [0,1)$, as we will explore in detail.

Finally, we emphasize that Eq.~\eqref{eq:KramerSIDTime} \deletetext{characterizes the probability of escape {\em at a given time}, rather than the mean escape time}\newtext{holds in probability}. As such, it does not constitute an Arrhenius-type law, which would describe the expected time to \newtext{reach a specified position} (this is also noted in the Appendix of~\cite{Aleksian2024}). The quantification of an Arrhenius law for the feedback OU process, and its relation to the underlying large deviation principles, is one of the main results of our work.


\section{First-passage Dynamics and Weak-Noise Large Deviations}
\label{sec:Weak}

Given the Markovian structure of the dynamics \newtext{of the process $(Z_t)_{t\geq0}$} in Eq.~\eqref{eq:ProcessMarkov}, the associated time-dependent infinitesimal generator takes the form:
\begin{equation}
\label{eq:InfinitesimalGenMarkovFeedback}
L_t = A_t(\beta) z \cdot \nabla + \frac{\epsilon}{2} \frac{\partial^2}{\partial x^2} \, ,
\end{equation}
where $z \coloneqq (x, \bar{x})$ and $\nabla$ denotes the gradient with respect to the variables $(x, \bar{x})$. \newtext{In principle,} this generator can be used to extract information on the \deletetext{escape}\newtext{first-passage} dynamics, in particular: (i) the \deletetext{escape}\newtext{first-passage} time \deletetext{distribution $f_T(x_f)$}\newtext{probability density}, and (ii) the mean \deletetext{escape}\newtext{first-passage} time\deletetext{$\bar{T}$, defined as the first moment of $f_T(x_f)$}.

\newtext{For the non-homogeneous Markov process with time-dependent infinitesimal generator given in~\eqref{eq:InfinitesimalGenMarkovFeedback}, the first-passage time is defined as
\begin{equation}
    T_s(z_s) = \inf \left\lbrace t - s \geq 0 : X_t > x_f  \right\rbrace \, ,
\end{equation}
with initial condition $Z_s=z_s \coloneqq (x_s,\bar{x}_s)$ at time $s$. Its average, the mean first-passage time, is given by 
\begin{equation}
    \label{eq:MeanFPTFeedback}
    \bar{T}_s(z_s) = \mathbb{E}_{s,z_s} \left[ T - s \right] = \mathbb{E} \left[ \inf \left\lbrace t - s \geq 0 : X_t > x_f \right\rbrace | Z_s = z_s   \right] \, ,
\end{equation} 
which is the expected time (from time $s$) it takes for the process to hit $x_f$. Notice that both the first-passage time and its average depend on the starting time $s$ and starting position $z_s$ (due to the non-homogeneity of the process). 

}

\newtext{In order to describe a general procedure to calculate the first-passage time probability density and its first moment $\bar{T}_s(z_s)$, we introduce the transition probability density of the process $\left( Z_t \right)_{t \geq 0}$, with absorbing boundary at $x_f$, which is written as
\begin{equation}
    \label{eq:TransProbaDensMarkovNonHomo}
    \rho(z,t|z_s,s) \coloneqq P(Z_t = z|Z_s=z_s) \, ,
\end{equation}
for $t \geq s$. Given the pointwise initial condition, the transition probability density above is already normalized.
We remark that $P(Z_t = z|Z_s=z_s)$ is shorthand for the probability that $Z_t$ lies in a small interval $[x, x + dx) \times [\bar{x}, \bar{x} + d\bar{x})$ conditioned on having started the process at $z_s$ at time $s$, so that the probability density satisfies $\mathbb{P}(Z_t = z|Z_s = z_s) = P(Z_t = z| Z_s=z_s) \, dx d\bar{x}$, where $\mathbb{P}$ is the probability distribution.

\newtext{Returning to the transition probability density function $\rho(z,t|z_s,s)$ in~\eqref{eq:TransProbaDensMarkovNonHomo}, we remark that it satisfies the following Fokker--Planck boundary value problem:
\begin{equation}
\label{eq:ForwardTransitionProba}
\frac{\partial \rho(z,t|z_s,s)}{\partial t} = L_t^\dagger \rho(z,t|z_s,s) \, , \quad \textrm{with}\quad \rho(z,s|z_s,s) = \delta(z - z_s) \, \quad \textrm{and}\quad \rho((x_f,\bar{x}),t|z_s,s) = 0 \;\; \forall \bar{x}\in \mathbb{R} \, ,
\end{equation}}
with $t \geq s$ and where $L_t^\dagger$ denotes the adjoint of the generator $L_t$ in Eq.~\eqref{eq:InfinitesimalGenMarkovFeedback}. We also have that} the \deletetext{escape}\newtext{first-passage} time \deletetext{distribution}\newtext{probability density} is given by
\newtext{\begin{equation}
\label{eq:FirstPassageTimeDistr}
f_{s,t}(z_s) \coloneqq - \frac{\partial Q_{s,t}(z_s)}{\partial t} \, ,
\end{equation}
}
where
\newtext{
\begin{equation}
    \label{eq:Survival}
    Q_{s,t}(z_s) \coloneqq \int_{-\infty}^{x_f} \int_{-\infty}^{\infty} \rho(z,t|z_s,s) \, d\bar{x} \, dx ,
\end{equation}
is the survival probability density function, which is the probability that the particle starting at $z_s$ at time $s$ has {\em not hit} $x_f$ by time $t \geq s$ (notice that the variable $\bar{x}$ is also integrated out).}

\newtext{Therefore,} the mean \deletetext{escape}\newtext{first-passage} time $\bar{T}_s(z_s)$ satisfies the following boundary value problem:
\newtext{\begin{equation}
\label{eq:BackwardMeanEscapeTime}
\begin{cases}
    L_s \, \bar{T}_s(z_s) = -1, 
    & \quad \text{for } x_s < x_f , \, \bar{x}_s < x_f \\[8pt]
    \bar{T}_s(z_s) = 0, 
    & \quad \text{for } x_s = x_f, \, \bar{x}_s<x_f \, ,
\end{cases}
\end{equation}
whose derivation is given in Appendix~\ref{ref:appa}}. \newtext{Because the} infinitesimal generator $L_t$ is explicitly time-dependent due to the drift term, and the diffusion is degenerate, since no noise acts directly on the dynamics of $\bar{X}_t$, exact analytical solutions to Eqs.~\eqref{eq:ForwardTransitionProba} and~\eqref{eq:BackwardMeanEscapeTime} are generally not available. 

However, asymptotic approaches are possible. For instance, approximate analytical solutions for both the autonomous and non-autonomous one-dimensional Ornstein--Uhlenbeck processes and for stochastic resonance have recently been obtained using matched asymptotic expansions~\cite{Moon2020, Moon2021, Giorgini2020, Giorgini2024}. While this approach could in principle be extended to the two-dimensional setting considered here, identifying the appropriate boundary layers in two dimensions is highly non-trivial and would require a separate in-depth study.

Instead, here we use the framework of weak-noise large deviation theory to derive asymptotic results for the \deletetext{escape}\newtext{first-passage} dynamics of the feedback OU process. In the weak-noise regime, we can analyze the asymptotic behavior of the probability \newtext{density $\rho(z,t|z_0,0)$} by solving an effective approximation of Eq.~\eqref{eq:ForwardTransitionProba}, that is the associated Hamilton--Jacobi equation.

In Section~\ref{subsec:WeakFrame}, we briefly review the general weak-noise large deviation framework for Markov processes—see~\cite{Freidlin1984,Touchette2009,Grafke2019} for more comprehensive treatments. We then illustrate how this framework can be used to extract relevant information about \deletetext{escape}\newtext{first-passage} dynamics and apply it to both the standard OU process (Section~\ref{subsec:WeakOU}) and the feedback OU process (Section~\ref{subsec:WeakFeedbackOU}).

\subsection{Weak-Noise Large Deviations}
\label{subsec:WeakFrame}

\newtext{For simplicity,} we introduce the weak-noise large deviation framework by considering \newtext{a general two-dimensional} Markov process, \newtext{denoted, with slight abuse of notation,} $(Z_t)_{t \geq 0}$. \newtext{We initially keep the notation as general as possible, since the framework extends to any $d$-dimensional Markov process, and then specialize to the case of interest in Sec.~\ref{subsec:WeakFeedbackOU}.} \newtext{The general process is} governed by the stochastic differential equation, 
\begin{equation}
\label{eq:DiffusionGeneral}
dZ_t = F_t(Z_t) \, dt + \sqrt{\epsilon} \, \sigma \, dW_t \, ,
\end{equation}
where $F_t$ is a (possibly time-dependent) $2$-dimensional drift vector field, $\sigma \in \mathbb{R}^{2 \times 2}$ is a constant noise matrix, $D = \sigma^T \sigma$ is the noise covariance (or diffusion) matrix, and $W_t$ denotes a $2$-dimensional Brownian motion. For simplicity, we assume additive noise, i.e., $\sigma$ does not depend on position, and we consider a pointwise initial condition $Z_0 = z_0$. 
\newtext{We are interested in events wherein the process $(Z_t)_{t \geq 0}$ exhibits a specific outcome---for instance, when the trajectory reaches a prescribed set $A \subset \mathbb{R}^2$ at time $T$, that is\ $Z_T \in A$. While such events are impossible in the deterministic case ($\epsilon = 0$), they typically occur in the presence of noise ($\epsilon > 0$), though of course with decreasing frequency as the noise intensity vanishes, $\epsilon \to 0$.}

\newtext{Similarly to the previous treatment, we introduce the normalized transition probability density function $\rho(z,t | z_0,0)$ as in Eq.\ \eqref{eq:TransProbaDensMarkovNonHomo} with absorbing boundary $A \subset \mathbb{R}^2$. This function satisfies a Fokker--Planck boundary value problem equivalent to \eqref{eq:ForwardTransitionProba} and written below for thoroughness}
\deletetext{Similarly to Section \ref{sec:Weak}, we are interested in the asymptotic behavior of the probability density, }
\deletetext{\begin{equation}
\label{eq:ProbaDensityMarkov}
P_t(z) = \int \delta(z'-z'') \rho(z,t|z'',0) \, dz'' \, ,
\end{equation}
satisfies the following Fokker--Planck equation, which} 
\begin{equation}
\label{eq:FPGeneral}
\partial_t \rho(z,t|z_0,0) = - \nabla \cdot \left( F_t(z) \, \rho(z,t|z_0,0) \right) + \frac{\epsilon}{2} \nabla \cdot \left( D \, \nabla \rho(z,t|z_0,0) \right) \, ,
\end{equation}
with initial condition \newtext{$\rho(z,0|z_0,0) = \delta(z - z_0)$ and absorbing boundary $\rho(z,t|z_0,0)=0$ for $z \in A$ and $t \geq 0$}.

Now, by assuming an exponential scaling in $\epsilon$ \deletetext{and making}\newtext{we make} the following change of variable,
\begin{equation}
\label{eq:ChangeVar}
\rho(z,t|z_0,0) = \exp \left[- \epsilon^{-1} S_t(z) \right] \, ,
\end{equation}
\newtext{where, for parsimony of notation, on the right hand side\ the dependence on the initial time and point is implicit.} Taking the weak-noise limit, $\epsilon \rightarrow 0$, and retaining terms up to order $O(\epsilon^{-1})$, gives the following Hamilton--Jacobi equation~\cite{Grafke2019}:
\begin{equation}
\label{eq:HamiltonJacobi}
- \partial_t S_t(z) = H_t(z, \nabla S_t(z)) \, .
\end{equation}
Here, the time-dependent Hamiltonian is given by
\begin{equation}
\label{eq:Hamiltonian}
H_t(z, p)= \frac{1}{2} \, p \cdot D \, p + F_t(z) \cdot p \, ,
\end{equation}
where $p$ denotes a vector of adjoint variables, or momentum coordinates. The approximation introduced through the change of variable~\eqref{eq:ChangeVar} with $\epsilon \ll 1$ is commonly referred to as the WKBJ (after Wentzel, Kramers, Brillouin and Jeffreys) ansatz in many areas of physics, including stochastic processes~\cite{Assaf2017, Weber2017, Moon2020}.

The Hamilton--Jacobi equation~\eqref{eq:HamiltonJacobi} is a first-order partial differential equation whose solution can be obtained from the characteristic curves of the associated Hamiltonian dynamics, which are
\begin{equation}
\label{eq:HamiltonEqGen}
\dot{z}_t = \nabla_p H_t(z, p) \, \qquad \textrm{and} \qquad \dot{p}_t = - \nabla_z H_t(z, p) \, ,
\end{equation}
subject to appropriate boundary conditions. By integrating along these characteristic trajectories—often referred to as {\em instantons} in the large deviation literature~\cite{Grafke2019,Assaf2017}—the action functional in Eq.~\eqref{eq:HamiltonJacobi} takes the following form;
\begin{equation}
\label{eq:ActionExplicit}
S_t(z) = \int_0^t \left[p \cdot \dot{z} - H_t(z, p) \right] \, dt + S_0(z_0) \, .
\end{equation}
This expression reflects the fact that the action is the generating function of a canonical transformation, linking the first-order PDE~\eqref{eq:HamiltonJacobi} with the second-order Hamiltonian dynamics of Eq.~\eqref{eq:HamiltonEqGen}~\cite{Arnold1989,Goldstein2001}.

In the weak-noise regime, the action so obtained represents the dominant contribution  to the probability density function \newtext{$\rho(z,t|z_0,0)$} in Eq.~\eqref{eq:ChangeVar}, wherein, with a slight abuse of notation, we denote both the full action and its leading-order contribution at $O(\epsilon^{-1})$ with the same symbol. This action determines the asymptotic behavior in $\epsilon^{-1}$ of the probability of finding the process at \newtext{$z_f \in A$} for the first time $T$\deletetext{, as in Eq.~\eqref{eq:ProbaEscapeT}}, viz.\
\begin{equation}
\label{eq:ProbaEscapeWeakT}
\rho(z_f,T|z_0,0) \approx \exp \left[ - \epsilon^{-1} S_T(z_f) \right] \, .
\end{equation}
\newtext{At leading order in $\epsilon^{-1}$, this characterizes the survival probability density function.} \deletetext{Moreover}\newtext{Therefore}, upon taking the time derivative of Eq.~\eqref{eq:ProbaEscapeWeakT}, we find that in the weak-noise limit the \deletetext{escape}\newtext{first-passage} time distribution in Eq.~\eqref{eq:FirstPassageTimeDistr} is also approximated by \newtext{$\rho(z_f,T|z_0,0)$} of Eq.~\eqref{eq:ProbaEscapeWeakT}.

The optimal trajectory $z_t$ is that which minimizes the action, which defines the instanton.  Thus, the instanton corresponds to the most likely path connecting the initial point $z_0$ to the final point $z_T = z_f$ as $\epsilon \to 0$, thereby providing the dominant path through which the system reaches $z_f$ at time $T$.
By integrating over time, Eq.~\eqref{eq:ProbaEscapeWeakT} leads to the Kramers-type, or Arrhenius-type, formula as
\begin{equation}
\label{eq:ProbaEscapeWeak}
\mu(z_f|z_0,0) \approx \exp \left[ - \epsilon^{-1} U(z_f) \right] \, ,
\end{equation}
where \newtext{$\mu$ is the probability density on the state space induced by time integration and}
\begin{equation}
\label{eq:QuasiPotential}
U(z_f) \coloneqq \inf_{T \geq 0} S_T(z_f)
\end{equation}
is the quasi-potential, or the large deviation rate function. We interpret $U(z_f)$ as an effective energy barrier that the system must overcome to reach $z_f$, and it governs the leading-order behavior of the \deletetext{escape}\newtext{first-passage} probability. Additionally, the optimal \deletetext{escape}\newtext{first-passage} time can be determined as
\begin{equation}
\label{eq:TypicalEscapeWeak}
\tau(z_f) = \underset{T \geq 0}{\text{arginf}} \,\, S_T(z_f) \, .
\end{equation}
\newtext{We stress that the equalities in Eq.\ \eqref{eq:ProbaEscapeWeakT} and Eq.\ \eqref{eq:ProbaEscapeWeak} are exact only at the logarithmic scale. Consequently, the right hand side\ of both equations are not normalized, since terms of order $O(\epsilon^{-1})$ are neglected. This explains the apparent divergence obtained when integrating Eq.\ \eqref{eq:ProbaEscapeWeakT} over time.}

We note that in general this optimal first-passage time $\tau(z_f)$ is not equivalent to the mean first-passage time \newtext{$\bar{T}_0(z_0)$ defined in Eq.\ \eqref{eq:MeanFPTFeedback} and} approximated by the inverse Kramers rate\newtext{; $\bar{T}_0(z_0) \approx \exp \left[ \epsilon^{-1} \, U(z_f) \right]$}\deletetext{, which corresponds to the solution of the backward equation~\eqref{eq:BackwardMeanEscapeTime}}. Although the two quantities are related, they have distinct interpretations. The optimal first-passage time is \deletetext{based on the idea that the system drifts toward $z_f$ as slowly as possible to minimize the action. However, real \deletetext{escape}\newtext{first-passage} events are noise induced and occur asymptotically on a finite timescale given by the inverse Kramers rate in Eq.~\eqref{eq:ProbaEscapeWeak}.}\newtext{defined in the limit $\epsilon \to 0$, where it is given by the value of $T$ at which the action $S_T(z_f)$ reaches its minimum. It is therefore independent of $\epsilon$ and should not be confused with either the mean or the typical first-passage time with finite noise. Instead, it captures the intrinsic tendency of the process in the absence of noise. For example, in the Ornstein–Uhlenbeck process of Sec.\ \ref{subsec:WeakOU} below, the minimum of the action is attained only as $T \to \infty$, reflecting the fact that the noiseless dynamics keeps the process at the fixed point indefinitely (before escaping). In contrast, real first-passage events are noise-induced and occur asymptotically on the finite timescale set by the inverse Kramers rate in Eq.~\eqref{eq:ProbaEscapeWeak}.}

\subsection{The Weak-Noise Regime of the Ornstein--Uhlenbeck Process}
\label{subsec:WeakOU}


As noted above, a canonical analytically tractable stochastic process in statistical physics is the Ornstein--Uhlenbeck process, which
is obtained by removing the feedback term, $\frac{1}{t} \int_0^t X_s \, ds$, from Eq.~\eqref{eq:Process}. \newtext{The result derived in the following can be found in the standard literature of stochastic processes, see for instance~\cite{Risken1996,Pavliotis2014}.} The transition probability density function is a Gaussian that can be computed explicitly by solving the Fokker--Planck equation
\begin{equation}
\label{eq:FPOU}
\partial_t \rho(x,t|0,0) = L^\dagger \rho(x,t|0,0) \, ,
\end{equation}
with initial condition $\rho(x,0|0,0) = \delta(x)$, where $L$ is the infinitesimal Markov generator given by
\begin{equation}
\label{eq:OUGen}
L = - x \frac{d}{dx} + \frac{\epsilon}{2} \frac{d^2}{dx^2} \, .
\end{equation}
Equation \eqref{eq:FPOU} can be solved using Fourier transform or heat kernel methods giving the transition probability density as
\begin{equation}
\label{eq:PropaOU}
\rho(x,t|0,0) = \sqrt{\frac{1}{\pi \epsilon (1 - e^{-2t})}} \exp \left[ - \frac{1}{\epsilon} \frac{x^2}{(1 - e^{-2t})} \right] \, ,
\end{equation}
which is the probability of finding the process at position $x$ at time $t$\newtext{, starting at $x=0$ at $t=0$}.

Because Eq.~\eqref{eq:PropaOU} takes a large deviation form, we identify the action $S^{\text{OU}}_T$ in Eq.~\eqref{eq:ProbaEscapeWeakT} as
\begin{equation}
\label{eq:ActionOU}
S^{\text{OU}}_T(x_f) \coloneqq \frac{x_f^2}{1 - e^{-2T}} \, .
\end{equation}
As discussed in Section \ref{subsec:WeakFrame}, this gives an asymptotic characterization of the probability of finding the OU process at $x_f$ for the first time $T$; it provides the leading-order asymptotic form of the \deletetext{escape}\newtext{first-passage} time distribution in Eq.~\eqref{eq:FirstPassageTimeDistr}. By solving Eqs.~\eqref{eq:HamiltonEqGen} for the Hamiltonian in Eq.~\eqref{eq:Hamiltonian}, which in this case is $H(x, p) = \frac{1}{2}p^2 - p x$, we obtain the instanton equation
\begin{equation}
\label{eq:InstantonOU}
x^{\text{OU}}_t = x_f \frac{\sinh t}{\sinh T} \, ,
\end{equation}
which gives the most probable path from $x^{\text{OU}}_0 = 0$ to $x^{\text{OU}}_T = x_f$. Integrating Eq.~\eqref{eq:PropaOU} in time yields the probability of being at $x$ as an implicit function of time;
\begin{equation}
\label{eq:ProbaEscapeOU}
\mu(x|0,0) = \frac{|x|}{\sqrt{\pi} \epsilon} \int_{\frac{|x|}{\sqrt{\epsilon}}}^\infty \frac{e^{-t^2}}{t^2 - \frac{|x|}{\sqrt{\epsilon}}} \, dt \, ,
\end{equation}
and its leading-order behavior in $\epsilon^{-1}$, which can also be obtained from Eq.~\eqref{eq:QuasiPotential}, is given by the following quasi-potential
\begin{equation}
\label{eq:QuasiPotentialOU}
U^{\text{OU}}(x_f) = x_f^2 \, ,
\end{equation}
which is quadratic in the final position as expected. Moreover, we note that the quasi-potential for the OU process coincides with $H(x_f)$ in Eq.~\eqref{eq:KramerSIDExponent}.

Finally, the \deletetext{escape}\newtext{first-passage} time of the process generally depends on the noise intensity $\epsilon$. Although the mean \deletetext{escape}\newtext{first-passage} time is finite and given by the inverse Kramers rate, in the weak-noise limit inspection of Eq.~\eqref{eq:TypicalEscapeWeak} shows that the optimal \deletetext{escape}\newtext{first-passage} time diverges; $\tau \to \infty$. 
Therefore, the weak-noise OU process tends to remain near the bottom of the potential well for as long as possible in order to minimize the action, eventually escaping the quadratic trap through a single large fluctuation.

\subsection{The Weak-Noise Regime of Feedback Ornstein--Uhlenbeck Processes}
\label{subsec:WeakFeedbackOU}

We now turn our attention to the feedback OU process. Rather than analyzing the non-Markovian process $X_t$ defined by Eq.~\eqref{eq:Process}, we consider the two-dimensional Markovian process $Z_t = (X_t, \bar{X}_t)$ defined by Eq.~\eqref{eq:ProcessMarkov}. \newtext{Using the notation of }\deletetext{The dynamics is governed by} Eq.~\eqref{eq:DiffusionGeneral}, \deletetext{with a}\newtext{the} deterministic time-dependent drift \newtext{is} given by
\begin{equation}
\label{eq:MarkovDriftFeedback}
F_t(Z_t) = A_t(\beta) Z_t \, ,
\end{equation}
and a noise matrix
\begin{equation}
\label{eq:NoiseMatrix}
\sigma = D = 
\begin{pmatrix}
1 & 0 \\
0 & 0 
\end{pmatrix} \, ,
\end{equation}
showing that noise acts only on the $X_t$ component. (We discuss the consequences of the structure of the noise matrix below.) We consider the initial condition $Z_0 = (x_0,\bar{x}_0) = (0,0)$, which also regularizes the singularity at $t = 0$, and we aim to study the \deletetext{escape}\newtext{first-passage} dynamics through $x_f > 0$.


The time-dependent Hamiltonian is given by
\begin{equation}
\label{eq:HamiltonianFeedbackOU}
H_t(z,p) = \frac{p_x^2}{2} - \left( x -  \beta \bar{x} \right) p_x + \frac{(x - \bar{x})}{t} p_{\bar{x}} \, ,
\end{equation}
where $z = (x, \bar{x})$ and $p = (p_x, p_{\bar{x}})$.
The solutions to the Hamilton--Jacobi equation~\eqref{eq:HamiltonJacobi} are obtained by solving the associated Hamilton equations~\eqref{eq:HamiltonEqGen} and are:
\begin{equation}
\label{eq:HamiltonEqBoundarySpec}
\begin{cases}
\dot{x} = p_{x} - x + \beta \bar{x} \\
\dot{\bar{x}} = \dfrac{x - \bar{x}}{t} \\
\dot{p}_{x} = p_{x} - \dfrac{p_{\bar{x}}}{t} \\
\dot{p}_{\bar{x}} = - \beta p_{x} + \dfrac{p_{\bar{x}}}{t}
\end{cases}
\qquad
\begin{cases}
x_0 = 0 \\
\bar{x}_0 = 0 \\
p_{x,T} = \lambda \\
p_{\bar{x},T} = 0
\end{cases} \, .
\end{equation}
In Eq.\ \eqref{eq:HamiltonEqBoundarySpec}, $\lambda$ is a parameter implicitly determined by the final condition $x_T = x_f$ through the Hamiltonian dynamics. As we highlight in the derivation below, it is more convenient to impose final conditions on the adjoint variables rather than initial conditions on the adjoint variables or final conditions on the coordinate variables. 
We also refer the reader to Section 2.4.1 of~\cite{Schulz2006} for an explanation of how to transform final conditions on position variables into final conditions on adjoint variables within a Lagrangian framework.

\newtext{We note that, in principle, for non-autonomous driving, the choice of boundary conditions in the propagator (see \ Eq.\ \eqref{eq:FPGeneral}) can influence even the leading-order behavior in the large-deviation regime ($O(\epsilon^{-1})$), because first-passage events may depend on whether earlier crossings are excluded. This would in turn modify the boundary conditions in~\eqref{eq:HamiltonEqBoundarySpec}. 
However, in the feedback OU process considered here the boundary conditions do not affect the rate function. 
The process starts at the stable fixed point \((x,\bar{x})=(0,0)\), and we ask for a crossing at a distant value \(x_f > 0\). In the weak-noise limit the deterministic drift drives trajectories into the fixed point where they vanish, so waiting before departure incurs no cost in the action. 
As a result, the minimization problem leading to the instanton is the same with or without a ``no earlier crossing'' constraint: the optimal path leaves the fixed point once, at the time that minimizes the action, and reaches \(x_f\) at \(T\).}


Starting from Eqs.~\eqref{eq:HamiltonEqBoundarySpec}, we derive a simplified differential equation the solution to which gives the instanton. 
Importantly, the adjoint equations are decoupled from the coordinate variables, which allows us to solve them independently and substitute their solution into the coordinate equations.
We begin with the equation for $\dot{p}_{x}$. \deletetext{To simplify the analysis, we assume that $p_{\bar{x}}$ does not explicitly depend on $p_{x}$. Although this assumption may conflict with the last equation in~\eqref{eq:HamiltonEqBoundarySpec}, we adopt it here as an \textit{ansatz} and will post hoc verify its validity.} \newtext{The formal solution is obtained by treating $p_{\bar{x}}$ as a source term,} and using separation of variables and variation of constants, \deletetext{we obtain the following}\newtext{and is given by}
\begin{equation}
\label{eq:pxt}
p_{x,t} = \lambda e^{- (T - t)} - \int_t^T \frac{p_{\bar{x},s} e^{(t - s)}}{s} \, ds \, ,
\end{equation}
where the constant of integration is fixed by the boundary condition $p_{x,T} = \lambda$. Note that, because the momentum grows exponentially in time, this is why it is preferable to impose a final rather than an initial condition on the momentum, which naturally constrains the unbounded growth of the solution.


Substituting Eq.~\eqref{eq:pxt} into the equation for $p_{\bar{x}}$ and differentiating both sides with respect to $t$, we obtain the following second-order differential equation:
\begin{equation}
\label{eq:Forpyt}
\ddot{p}_{\bar{x}} = \dot{p}_{\bar{x}} \left( 1 + \frac{1}{t} \right) + \frac{p_{\bar{x}}}{t} \left( \beta - 1 - \frac{1}{t} \right) \, .
\end{equation}
This equation must be solved subject to the boundary conditions that $p_{\bar{x},T} = 0$, as in Eq.~\eqref{eq:HamiltonEqBoundarySpec}, and $\dot{p}_{\bar{x},T} = -\beta \lambda$, which follows directly from the evolution equation for $p_{\bar{x}}$.
The general solution of Eq.~\eqref{eq:Forpyt} is given in terms of special functions as
\begin{equation}
\label{eq:pyt}
p_{\bar{x},t} = t \left( c_1 \, \text{HypU} \left[ \beta, 1, t \right] + c_2 \, \text{LagL} \left[ -\beta, t \right] \right) \, ,
\end{equation}
where $c_1$ and $c_2$ are constants, the first term on the right-hand side is the Tricomi confluent hypergeometric function,
\begin{equation}
\label{eq:TricomiHU}
\text{HypU}[a,b,t] = \frac{1}{\Gamma(a)} \int_0^\infty e^{-st} s^{a-1} (1+s)^{b-a-1} \, ds \, ,
\end{equation}
and $\text{LagL}[n,t]$ is the Laguerre \deletetext{polynomial}\newtext{function}, defined as the solution of Laguerre's differential equation,
\begin{equation}
\label{eq:LagPolyn}
t y'' + (1 - t) y' + n y = 0 \, ,
\end{equation}
\newtext{is non-singular at $t=0$}\deletetext{where $n$ is a non-negative integer}. It can be shown that Eq.~\eqref{eq:pyt} satisfies $p_{\bar{x},0} = 0$, ensuring regularity in the limit $t \rightarrow 0$ in Eq.~\eqref{eq:HamiltonEqBoundarySpec}. The same regularity is obtained by formally considering the asymptotic limit of $\beta\rightarrow0$ or for $t\ll1$ or both. In both cases, $\dot{p}_{\bar{x}} = - \beta p_{x} + \dfrac{p_{\bar{x}}}{t}$ becomes $\dot{p}_{\bar{x}} = \dfrac{p_{\bar{x}}}{t}$, so that $p_{\bar{x}}=C t$ and hence $\dot{p}_{x} = p_{x} - C$\deletetext{, which in turn shows the validity of the ansatz considered}.

Imposing the two boundary conditions above determines the constants in Eq.~\eqref{eq:pyt} as
\begin{eqnarray}
\label{eq:c1}
c_1 &=& \frac{\beta \lambda \, \text{LagL} \left[ -\beta, T \right]}{ \beta T \, \text{HypU} \left[ 1 + \beta , 2, T \right] \text{LagL} \left[ -\beta, T \right] - T \, \text{HypU} \left[ \beta, 1, T \right] \text{LagL} \left[ -1 - \beta, 1, T \right]} \qquad \textrm{and} \\
\label{eq:c2}
c_2 &=& - \frac{\beta \lambda \, \text{HypU} \left[ \beta, 1, T \right]}{ \beta T \, \text{HypU} \left[ 1 + \beta , 2, T \right] \text{LagL} \left[ -\beta, T \right] - T \, \text{HypU} \left[ \beta, 1, T \right] \text{LagL} \left[ -1 - \beta, 1, T \right]} \, .
\end{eqnarray}
This completes the solution Eq.~\eqref{eq:pyt} of Eq.~\eqref{eq:Forpyt}, thereby providing an explicit expression for $p_{x,t}$ in Eq.~\eqref{eq:pxt}.

Differentiating the first equation in~\eqref{eq:HamiltonEqBoundarySpec}, substituting the expression for $\dot{\bar{x}}$, and using the first equation again to eliminate $\bar{x}$, we obtain the following second-order differential equation for $x_t$,
\begin{equation}
\label{eq:xtFin}
\ddot{x} = \dot{p}_x - \left( 1 + \frac{1}{t} \right) \dot{x} + \frac{p_x}{t} + (\beta - 1) \frac{x}{t} \, ,
\end{equation}
which can be rewritten, using Eq.~\eqref{eq:pxt}, as
\begin{equation}
\label{eq:xtFinEq}
\ddot{x} = \left( 1 + \frac{1}{t} \right) \left( p_x - \dot{x} \right) - \frac{1}{t} \left( p_{\bar{x}} - (\beta - 1) x \right) \, .
\end{equation}
This equation must be solved using $p_{x,t}$ from Eq.~\eqref{eq:pxt}, $p_{\bar{x},t}$ from Eq.~\eqref{eq:pyt}, and boundary conditions $x_0 = 0$ and $\dot{x}_0 = p_{x,0}$. The latter condition follows directly from the initial condition $\bar{x}_0 = 0$ and the Hamiltonian equation for $\dot{x}$. The dynamics of $\bar{x}_t$ can then be recovered by inverting the equation for $\dot{x}$ once the solution to Eq.~\eqref{eq:xtFinEq} is known.

In summary, we have reduced the full system Eqs.~\eqref{eq:HamiltonEqBoundarySpec} to a single second-order differential equation~\eqref{eq:xtFinEq}, whose solution provides the instanton $x_t$ connecting $x_0 = 0$ to $x_T = x_f$, which indirectly determines $\bar{x}_t$. The solution to Eq.~\eqref{eq:xtFinEq} must be computed numerically. We compare the instantons of the OU process and the feedback OU process in Fig.~\ref{fig1} \deletetext{and~\ref{fig2}} of Section~\ref{subsec:Instanton}.

Once the instanton solution is obtained, we can compute the action \newtext{using Eq.\ \eqref{eq:ActionExplicit}. Interestingly, by substituting the Hamiltonian \eqref{eq:HamiltonianFeedbackOU} along with Hamilton's equations \eqref{eq:HamiltonEqBoundarySpec} into Eq.\ \eqref{eq:ActionExplicit}, we obtain the following simplified form for the action:
\begin{equation}
    \label{eq:ActionSimplified}
    S_t(z) = \int_0^t \frac{p_{x,s}^2}{2} \, ds + S_0(z_0) \, ,
\end{equation}
with}\deletetext{using Eq.~\eqref{eq:ActionExplicit}, with $p = (p_x, p_{\bar{x}})$, the Hamiltonian~\eqref{eq:HamiltonianFeedbackOU}, and } $S_0(z_0) = 0$, since the process starts at the minimum of the potentials $V$ and $K$ given by Eqs.~\eqref{eq:SIDBackgroundFeedback} and~\eqref{eq:SIDInteractionFeedback}. The weak-noise approximation of the \deletetext{escape}\newtext{first-passage} probability through $x_f$ is then given by Eq.~\eqref{eq:ProbaEscapeWeak}, with:
\begin{equation}
\label{eq:FinalConditionz}
z_f = (x_f, \bar{x}_f) \, ,
\end{equation}
where $x_f$ is prescribed, and $\bar{x}_f$ is determined from the solution of Hamilton's equations. The quasi-potential and optimal \deletetext{escape}\newtext{first-passage} time are finally obtained using Eqs.~\eqref{eq:QuasiPotential} and~\eqref{eq:TypicalEscapeWeak}, respectively.

We conclude this section by highlighting the following. A natural approach to deriving the instanton equation would have been to consider the Lagrangian $L(z, \dot{z}) = \langle \dot{z} - F_t(z), D^{-1} (\dot{z} - F_t(z)) \rangle / 2$, where angle brackets denote the Euclidean scalar product.  This dual Lagrangian derivation would have ensured that we would have to grapple with consequences of the singular nature of the matrix $D$.
Indeed, the kernel of $D$ is one-dimensional and may encode important information about the dynamics. For instance, using the Moore--Penrose inverse\footnote{\newtext{Given any matrix \( A \in \mathbb{R}^{m \times n} \), its \emph{Moore--Penrose inverse} \( A^+ \in \mathbb{R}^{n \times m} \) is the unique matrix satisfying the following four conditions:
\text{(1)} $A A^+ A = A$,
\text{(2)} $A^+ A A^+ = A^+$,
\text{(3)} $(A A^+)^\top = A A^+$ and
\text{(4)} $(A^+ A)^\top = A^+ A$. In the special case that $A$ is square and invertible: \( A^+ = A^{-1} \). In the case discussed in the main text, $A \coloneqq D$ and $D^+ = D$.}} (pseudo-inverse), one would obtain an equation of the form $\ddot{x} = - (1 + t^{-1}) \dot{x} + t^{-1} (\beta - 1) x$, the solution of which is the instanton.  This equation for $\ddot{x}$, in contrast to Eq.~\eqref{eq:xtFinEq}, entirely misses the contribution of the adjoint dynamics. Thus, in such a formulation, it would be necessary to supplement the Euler--Lagrange equations with an additional kernel-dependent term, the analysis of which would require alternative methods. On the contrary, within the Hamiltonian framework no such issue arises: the degeneracy of $D$ is not a problem and the adjoint dynamics are fully incorporated into the solution.


\section{Numerical Results and Discussion}
\label{sec:Results}

\subsection{Instanton Dynamics}
\label{subsec:Instanton}

We now compare and contrast the classical and feedback OU processes allowing us to characterize the physical mechanisms responsible for \deletetext{escape}\newtext{first-passage dynamics}.  In particular, we present our results on the instanton dynamics leading the feedback OU process to \deletetext{escape}\newtext{first-passage} through $x_f > 0$, and compare them with the classical OU case. 

\begin{figure}[h!]
    \centering
    \includegraphics[width=\textwidth]{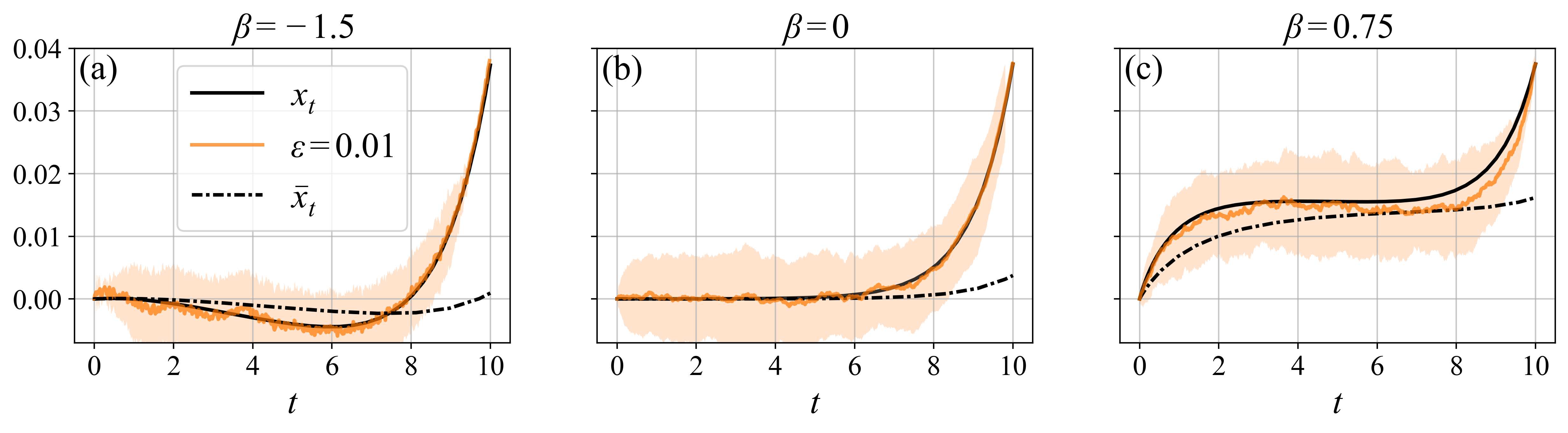}
    \caption{
    Comparison between the instanton $x_t$ solving Eq.~\eqref{eq:xtFinEq} for $T = 10$ and numerical simulations of the feedback OU process governed by Eqs.~\eqref{eq:ProcessMarkov}, for three values of the feedback parameter: $\beta < 0$ (a), $\beta = 0$ (b), and $\beta > 0$ (c). In the small-noise regime $\epsilon = 0.01$, the average simulated trajectory (solid orange line), computed over $10^7$ realizations and post-selected on reaching $x_f = 0.0375$ at time $T = 10$, closely follows the instanton. The shaded area shows one standard deviation around the average. The corresponding memory trajectory $\bar{x}_t$, obtained by inverting the equation for $\dot{\bar{x}}$ in Eq.~\eqref{eq:HamiltonEqBoundarySpec}, is also shown. These results support the accuracy of the analytical approximation. 
    }
    \label{fig1}
\end{figure}

To solve Eq.~\eqref{eq:xtFinEq} numerically we use the \texttt{solve\_ivp} solver from the \texttt{scipy.integrate} Python package, employing \texttt{RK45}, which corresponds to the explicit fourth order Runge--Kutta method of integration. In Fig.~\ref{fig1} we compare the numerical solution for the instanton $x_t$, obtained from the analytical approximation of Eq.~\eqref{eq:xtFinEq}, and the memory component $\bar{x}_t$, with simulations for three representative cases: $\beta < 0$ in Fig.~\ref{fig1}(a), $\beta = 0$ (OU process) in Fig.~\ref{fig1}(b), and $\beta > 0$ in Fig.~\ref{fig1}(c). Clearly, the average behavior of sample-path simulations conducted at finite but small noise $\epsilon$ compares well with the instanton $x_t$, thereby supporting\deletetext{the ansatz considered and, more generally,} our analytical derivation of Eq.~\eqref{eq:xtFinEq}.


Firstly, the case $\beta = 0$ shown Fig.~\ref{fig1}(b) exhibits the classical OU instanton $x_t^{\text{OU}}$ given by Eq.~\eqref{eq:InstantonOU}.
For completeness, we also show the memory $\bar{x}_t$, defined in Eq.~\eqref{eq:FeedbackDefinition}, which does not influence the dynamics in this case. As expected, the classical OU instanton corresponds to a monotonic exponential approach to $x_f$, as discussed in Section~\ref{subsec:WeakOU}.


Secondly, Fig.~\ref{fig1} reveals qualitatively distinct \deletetext{escape}\newtext{first-passage} mechanisms for negative and positive values of $\beta$, although both resemble the classical OU \deletetext{escape}\newtext{first-passage} at late times. For $\beta < 0$, Fig.~\ref{fig1}(a) shows that since it is detrimental to accumulate positive fluctuations toward $x_f$, the process instead accumulates fluctuations in the opposite direction. Because $\beta < 0$, this effectively moves the potential minimum toward $x_f$, storing a form of `potential energy' that is then released through a monotonic exponential approach to $x_f$. We refer to this behavior as the ‘slingshot’ effect.

For $\beta > 0$, Fig.~\ref{fig1}(c) shows that the process escapes optimally by initially accumulating small fluctuations toward $x_f$, which are stored in the memory $\bar{x}_t$, effectively shifting the minimum of the potential closer to $x_f$. As time progresses, the memory feedback becomes increasingly rigid due to averaging, which leads to a plateau in the intermediate time regime. However, since the effective potential minimum moves closer to $x_f$, \deletetext{escape}\newtext{first-passage} becomes less costly in terms of the action (see Fig.~\ref{fig3}) and we see the final exponential rise, again akin to the late time behavior of the classical OU case in Fig.~\ref{fig1}(b).



As previously noted, for simplicity we restrict our analysis to the case with $x_f > 0$, but note that the case $x_f < 0$ can be treated analogously. Indeed\deletetext{, as shown in Fig.~\ref{fig2}}, changing the sign of $x_f$ simply reflects the instanton trajectory, leaving the qualitative dynamics unchanged.
\deletetext{\begin{figure}[b]
    \centering
    \includegraphics[width=\textwidth]{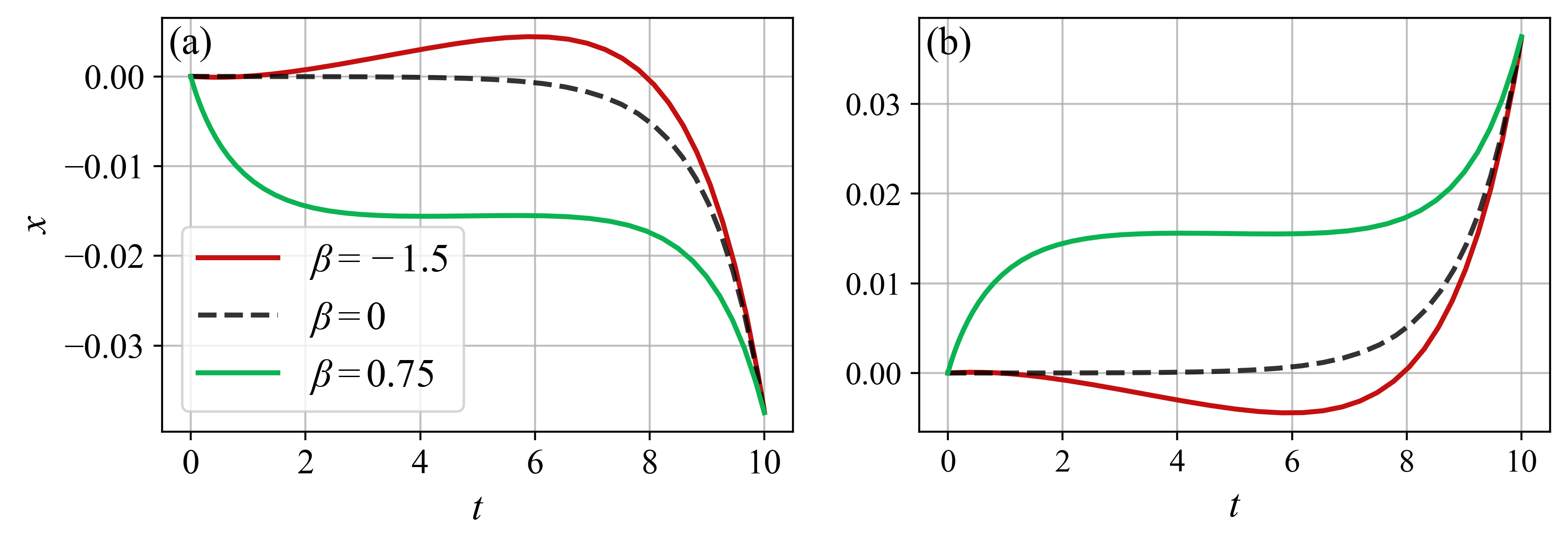}
    \caption{Instanton $x_t$ solutions of Eq.~\eqref{eq:xtFinEq} for $T=10$ and for the values of $\beta$ in Fig.~\ref{fig1}, plotted for a negative (a) and a positive (b) \deletetext{escape}\newtext{first-passage} point $x_f$. Changing the sign of $x_f$ corresponds to mirroring the instanton trajectories, thus fully characterizing the dynamics.}
    \label{fig2}
\end{figure}}
Having described the \deletetext{escape}\newtext{first-passage} mechanisms of the feedback OU process, next we analyze the action function $S_T(z_f)$ in order to assess the likelihood 
of these mechanisms.


\subsection{Time-Dependent Action}
\label{subsec:Action}

\begin{figure}[t]
    \centering
    \includegraphics[width=\textwidth]{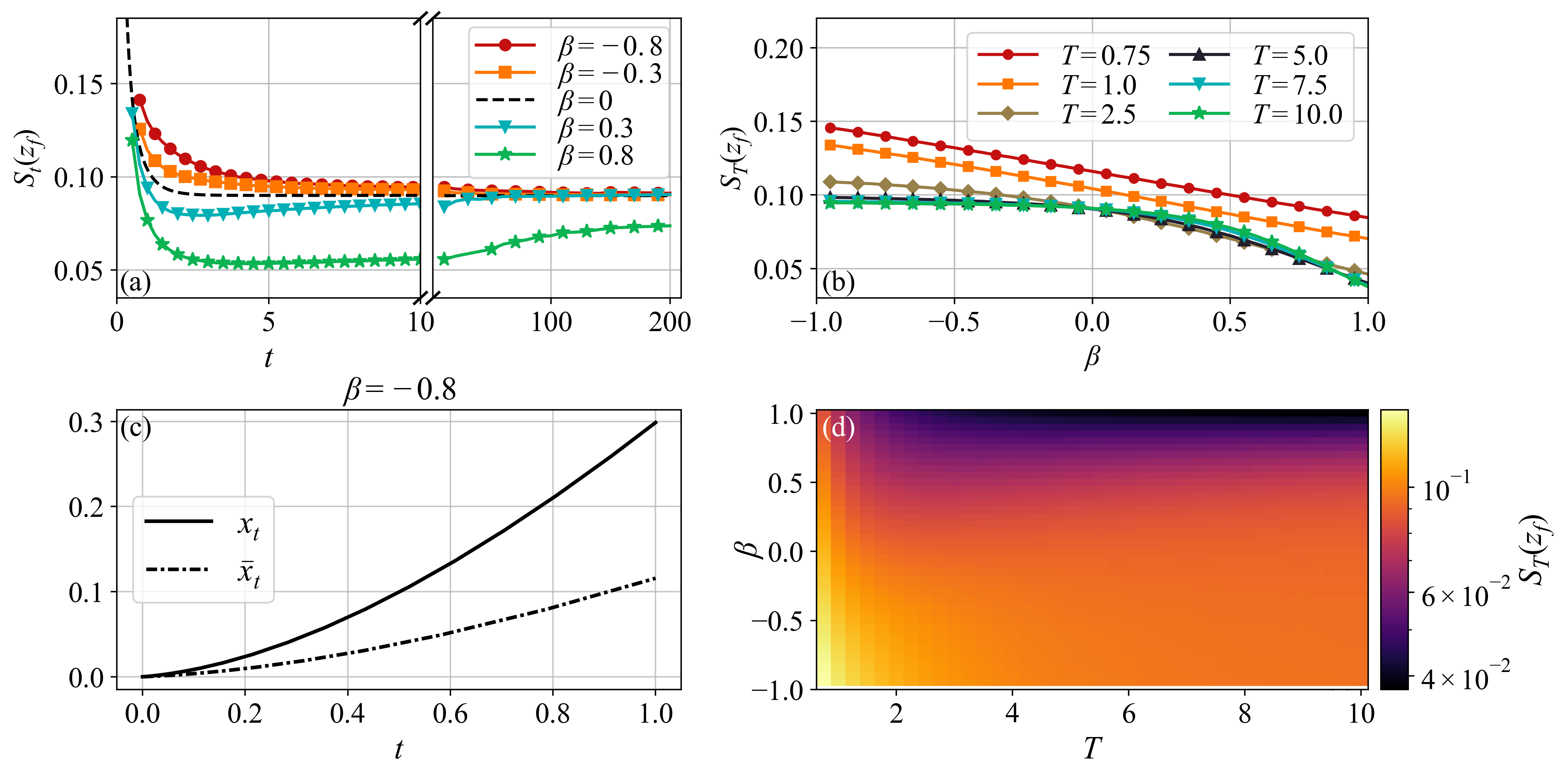}
    \caption{(a) The action $S_T(z_f)$ (Eq.~\ref{eq:ActionExplicit}) as a function of time $T$, at a fixed $x_f = 0.3$, for several values of $\beta$. The classical OU action is the black dashed line ($\beta = 0$), with the feedback OU process curves above ($\beta < 0$) and below ($\beta > 0$). (b) The action plotted as a function of $\beta$, at fixed $x_f = 0.3$, for several values of $T$. (c) The instanton $x_t$ and memory $\bar{x}_t$ for $\beta = -0.8$ and $T = 1.0$, showing nearly ballistic dynamics associated with the local minimum in (a) and the non-monotonic time behavior in (b). (d) A heatmap of the action plotted as a function of \deletetext{escape}\newtext{first-passage} time $T$ and feedback parameter $\beta$, at fixed $x_f = 0.3$.}
    \label{fig3}
\end{figure}

As defined in Eq.~\eqref{eq:ActionExplicit}, the action $S_T(z_f)$ depends both on the \deletetext{escape}\newtext{first-passage} point $z_f$ in Eq.~\eqref{eq:FinalConditionz} and on the \deletetext{escape}\newtext{first-passage} time $T$.  Moreover, as shown in Eq.~\eqref{eq:ProbaEscapeWeakT}, the action determines the probability of finding the process at $z_f$ for the first time $T$. In Fig.~\ref{fig3}, we collect a series of plots illustrating the important behavior of this action.


In Fig.~\ref{fig3}(a), we plot the action at a fixed \deletetext{escape}\newtext{first-passage} point $x_f = 0.3$ as a function of time $T$, for five values of $\beta$. When $\beta < 0$ ($\beta > 0$) the action lies above (below) the analytical action $S^{\text{OU}}_T$ of the classical OU process given in Eq.~\eqref{eq:ActionOU} and the black dashed line for $\beta=0$. This implies that, for a given $x_f$, the finite-time slingshot mechanism employed by the feedback OU process with $\beta < 0$ is less probable than the accumulation of positive fluctuations occurring for $\beta > 0$.  

\deletetext{When $\beta$ is sufficiently negative, Fig.~\ref{fig3}(a) shows the emergence of local and global minimum, with the local minimum occurring at short times and creating a sub-optimal time window for \deletetext{escape}\newtext{first-passage} dynamics.}\deletetext{Indeed}\newtext{For $\beta<0$, Fig.~\ref{fig3}(a) shows the emergence of a global minimum as $T \rightarrow \infty$}, as seen by the convergence of the colored curves towards the OU action from above. Thus, at large times, all \deletetext{escape}\newtext{first-passage} mechanisms asymptote to the standard OU dynamics, confirming the theoretical results on the long-time behavior of self-interacting diffusions discussed in Section~\ref{subsec:LongTime}. \newtext{On the other hand, at short times}, solving Hamilton's equations for \deletetext{this time and} $x_f = 0.3$ reveals that the instanton dynamics is nearly ballistic, as seen in Fig.~\ref{fig3}(c). 
\deletetext{However, as $\beta \rightarrow 0^-$ the sigmoidal structure, and hence the short-time local minimum, disappears as the global minimum is driven towards $T \rightarrow \infty$,}

In contrast, when $\beta > 0$ the global minimum of the action moves from $T \rightarrow \infty$ to a finite time, the value of which decreases as 
$\beta$ decreases. Therefore, accumulating and storing positive fluctuations via memory feedback is always more favorable than escaping through classical OU dynamics. This shift in the global minimum significantly influences the computation of the quasi-potential and optimal \deletetext{escape}\newtext{first-passage} time, as discussed in Section~\ref{subsec:Quasi-potential}.



As seen in Fig.~\ref{fig3}(b), for $\beta = 0$, classical OU \deletetext{escape}\newtext{first-passage} dynamics are more probable as $T$ increases, since the action decreases and approaches $\lim_{T \rightarrow \infty} S_T^{\text{OU}}(x_f)$. This \deletetext{continues for a range of negative $\beta$ until local minimum identified in Fig.~\ref{fig3}(a) becomes influential, as reflected in the $T = 1.0$ curve which crosses the others and disrupts the monotonicity of the action with escape time.}\newtext{monotonic behavior persists for all negative values of $\beta$. 
However, for $\beta > 0$, the behavior changes: reflecting the emergence of the global minimum shown in Fig.~\ref{fig3}(a), wherein the action is minimized at an intermediate time.}\deletetext{Thus, there is a threshold value $\beta\equiv\beta_c(x_f)$ such that for $\beta < \beta_c(x_f)$ ($\beta > \beta_c(x_f)$) the ballistic mechanism is more probable (less probable) than the slingshot mechanism.  We numerically estimate $\beta_c(x_f) \approx -0.35$, a value that remains approximately constant over the range $x_f \in [-0.45, 0.45]$, where numerics is reliable.}



\begin{figure}[t]
    \centering
    \includegraphics[width=\textwidth]{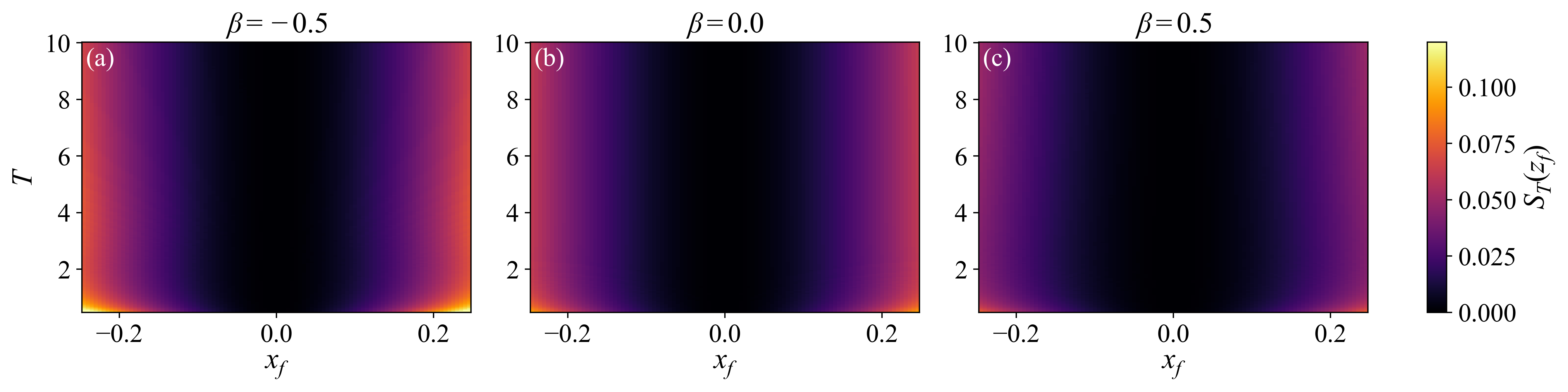}
    \caption{Heatmaps of the action $S_T(z_f)$ as a function of $x_f$ and $T$ for three feedback parameters: $\beta = -0.5$ (a), $\beta = 0$ (b), and $\beta = 0.5$ (c).}
    \label{fig4}
\end{figure}

A unified means of visualizing the information in Figs.~\ref{fig3}(a) and~(b) is obtained by constructing a heatmap of the action as a function of $\beta$ and \deletetext{escape}\newtext{first-passage} time $T$, at fixed $x_f = 0.3$, as shown in Fig.~\ref{fig3}(d).  Here we see that across all times the \deletetext{escape}\newtext{first-passage} dynamics for $\beta < 0$, be they  ballistic or slingshot, are less probable than for $\beta > 0$, where the accumulation of positive fluctuations dominates. 
Similarly, in Fig.~\ref{fig4} we plot heatmaps of the action as a function of $T$ and \deletetext{escape}\newtext{first-passage} point $x_f$, for three fixed values of the feedback parameter $\beta$. 
\deletetext{Firstly, }For all values of $\beta$ the figures show that escaping through boundary points farther from the origin is unlikely, and least likely for 
$\beta<0$. \deletetext{Secondly, the non-monotonicity discussed above is clearly seen in the patterns here, most notably for $\beta = -0.5$ in Fig.~\ref{fig4}(a), where \deletetext{escape}\newtext{first-passage dynamics} is less likely for $T = 5.0$ than for shorter or longer times.  This reflects the local minimum in the action seen in Fig.~\ref{fig3}(a) for $\beta < \beta_c(x_f)$, where short-time ballistic escape become the dominant mechanism.}




In Section~\ref{subsec:Quasi-potential}, we turn to the time-independent properties of the \deletetext{escape}\newtext{first-passage} dynamics, captured by the quasi-potential in Eq.~\eqref{eq:QuasiPotential}. Additionally, we examine the Kramers \deletetext{escape}\newtext{first-passage} rate and the optimal \deletetext{escape}\newtext{first-passage} time introduced in Eq.~\eqref{eq:TypicalEscapeWeak}.


\subsection{Quasi-Potential and Optimal \deletetext{escape}\newtext{First-Passage} Time}
\label{subsec:Quasi-potential}

The quasi-potential $U(z_f)$ is the large deviation rate function governing the probability of finding the process at a point $x_f$ in the small-noise limit $\epsilon \to 0$. It characterizes the effective `energy barrier' that the process must overcome to \deletetext{escape}\newtext{first-passage} through $x_f$. The larger the quasi-potential, the greater the barrier required for \deletetext{escape}\newtext{first-passage} and the longer the mean \deletetext{escape}\newtext{first-passage} time, which is approximated by the inverse of the probability in the Kramers rate formula in Eq.~\eqref{eq:ProbaEscapeWeak}.

\begin{figure}[h!]
    \centering
    \begin{minipage}{0.49\textwidth}
        \centering
        \includegraphics[width=\textwidth]{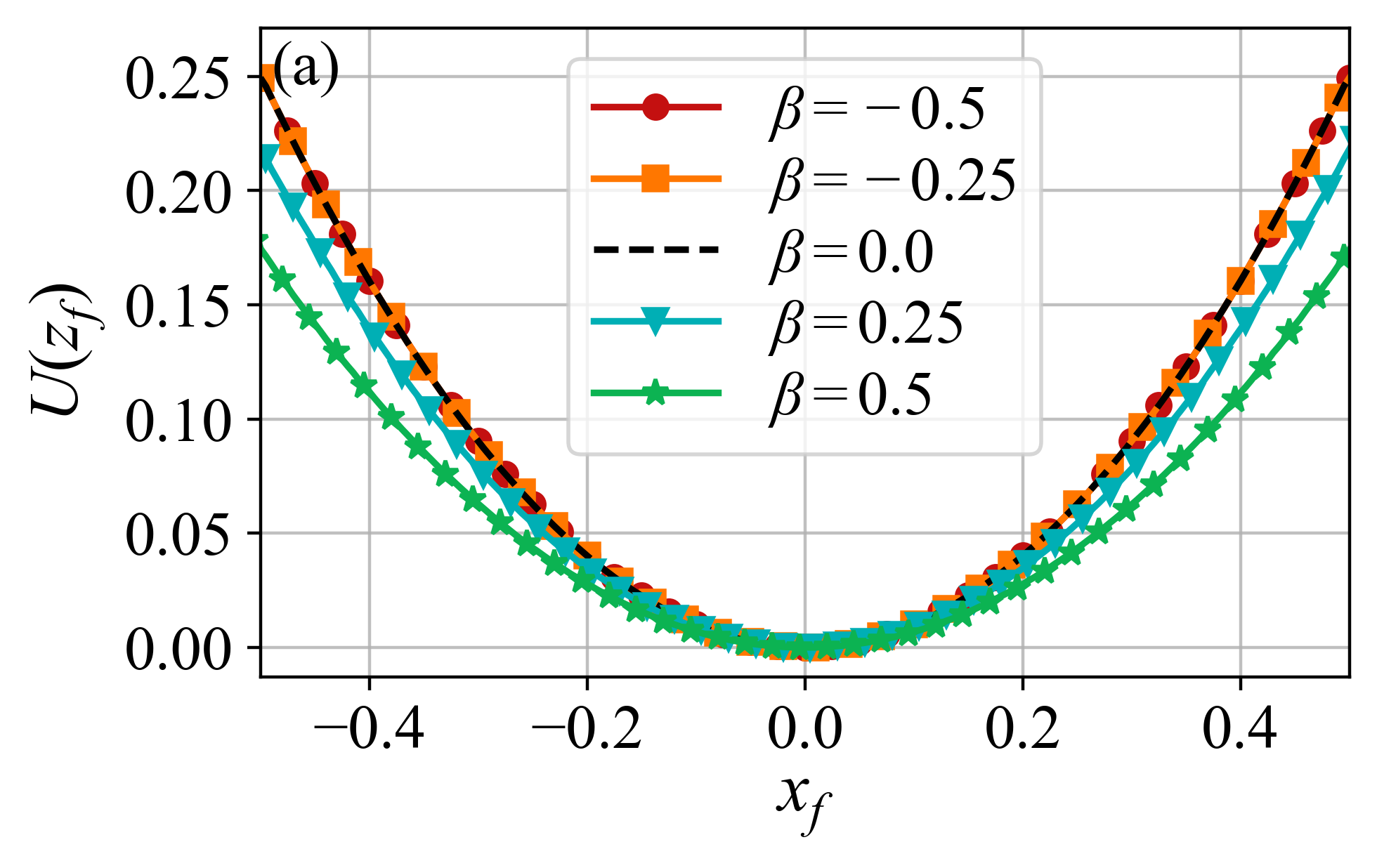}
    \end{minipage}
    \begin{minipage}{0.49\textwidth}
        \centering
        \includegraphics[width=\textwidth]{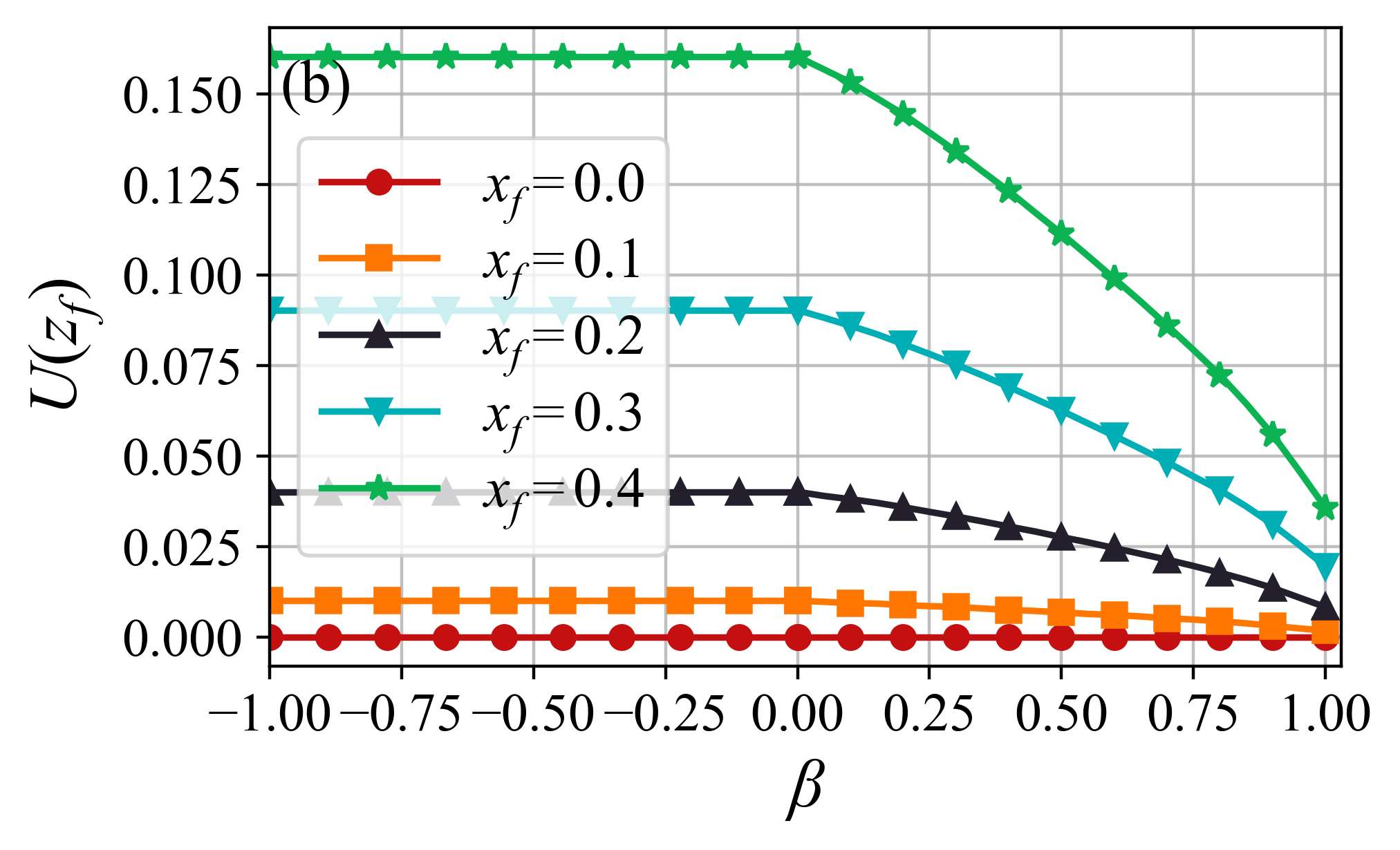}
    \end{minipage}\\[1em]
    \begin{minipage}{0.6\textwidth}
        \centering
        \includegraphics[width=\textwidth]{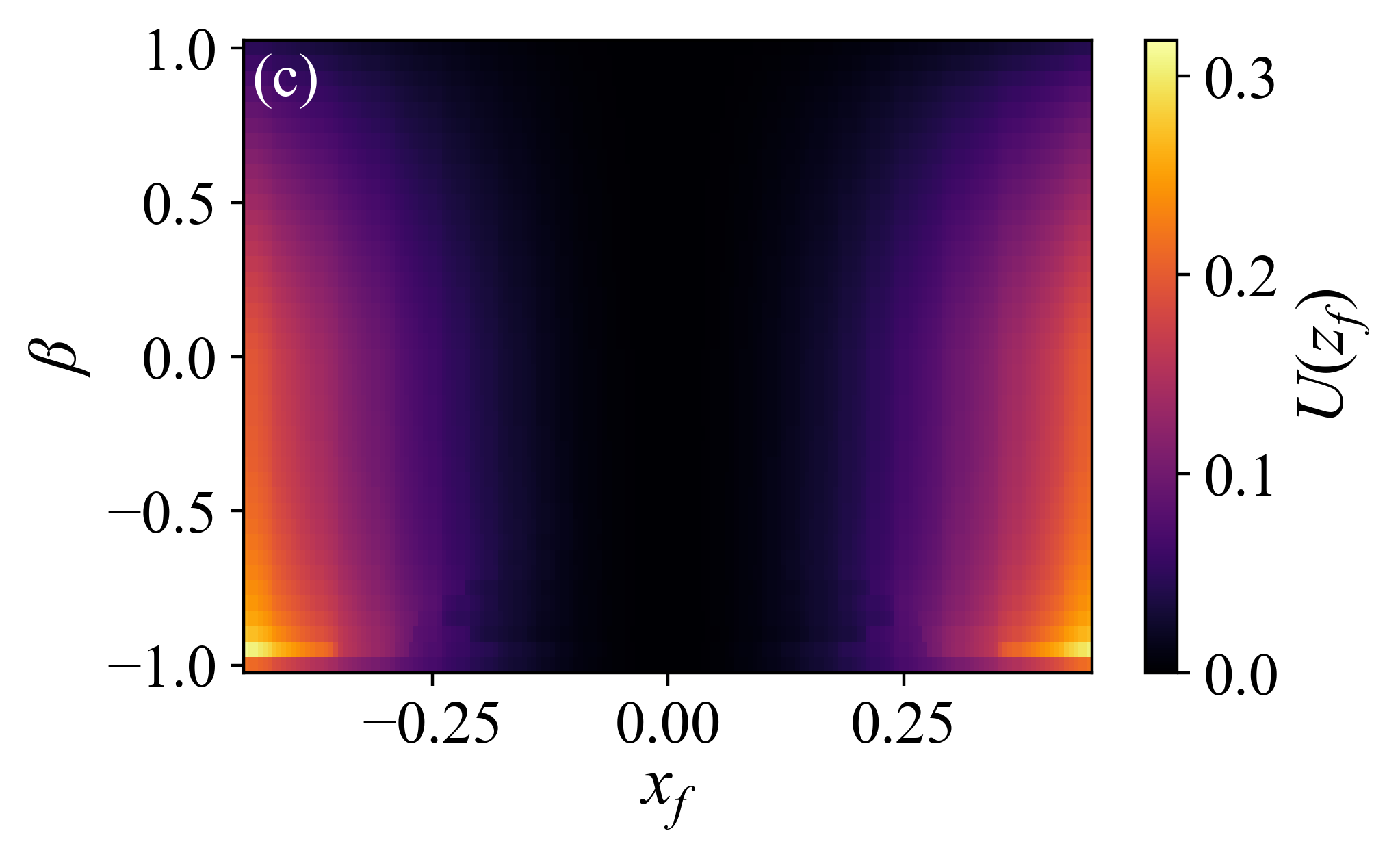}
    \end{minipage}
    \caption{(a) Quasi-potential $U(z_f)$ as a function of $x_f$. Curves for $\beta > 0$ lie below the OU result $U^{\text{OU}}(x_f)$, indicating that the most likely mechanism is the accumulation of positive fluctuations. For $\beta < 0$, curves collapse over $U^{\text{OU}}(x_f)$. (b) Quasi-potential $U(z_f)$ as a function of $\beta$ for several values of $x_f$. For $\beta > 0$, the curves decrease monotonically, reinforcing the enhanced likelihood of \deletetext{escape}\newtext{first-passage} via memory accumulation. For $\beta < 0$, the quasi-potential plateaus at $U^{\text{OU}}(x_f)$. (c) Heatmap of quasi-potential $U(z_f)$ as a function of $x_f$ and $\beta$.}
    \label{fig6}
\end{figure}

Fig.~\ref{fig6}(a) shows $U(z_f)$ as a function of the \deletetext{escape}\newtext{first-passage} point $x_f$ for five values of $\beta$ straddling the origin.  Clearly, for $\beta > 0$ the quasi-potential lies below the $\beta = 0$ black dashed line corresponding to the OU result $U^{\text{OU}}(x_f) = x_f^2$ given in Eq.~\eqref{eq:QuasiPotentialOU}. This is a consequence of the global minimum of the action $S_T(z_f)$ occurring at finite time shown in Fig.~\ref{fig3}(a). Thus, the most likely mechanism of reaching a given \deletetext{escape}\newtext{first-passage} point $x_f$ must have $\beta > 0$, and involves the accumulation of positive fluctuations in the memory. Moreover, as $\beta$ increases, this mechanism becomes more likely.


Importantly, this behavior is significantly different from the scenario in which the empirical occupation measure stabilizes before \deletetext{escape}\newtext{first-passage}, as discussed at the end of Section~\ref{subsec:LongTime} and reflected in the $\beta$-independent rate function in Eq.~\eqref{eq:KramerSIDExponent}. We argue that this difference arises because the quasi-potential in Eq.~\eqref{eq:QuasiPotential}, and shown in Fig.~\ref{fig6}(a), is the infimum of the action, rather than its value in the infinite-time limit, which would be a sufficient condition for the stabilization of the occupation measure.


As expected, the quasipotential coincides with the OU result when $\beta<0$, since the action $S_T$ reaches a global minimum as $T \rightarrow \infty$, as discussed in Fig.~\ref{fig3}.
In Fig.~\ref{fig6}(b), we plot the quasi-potential as a function of $\beta$ for five \deletetext{escape}\newtext{first-passage} points $x_f$. The plateaus for $\beta < 0$ correspond to the OU value of Eq.~\eqref{eq:QuasiPotentialOU}.  For $\beta > 0$, $U(z_f)$ decreases monotonically with $\beta$, reflecting the shift of the global minimum of the action to finite times. This reduction in the quasi-potential effectively lowers the escape barrier, thereby accelerating \deletetext{escape}\newtext{first-passage} dynamics by reducing the mean \deletetext{escape}\newtext{first-passage} time, which is approximated by taking the inverse of the Kramers rate in Eq.~\eqref{eq:ProbaEscapeWeak}. 

As previously, a heatmap perspective combines the behavior in a single figure, in this case the quasi-potential as a function of both $\beta$ and $x_f$, as shown in Fig.~\ref{fig6}(c).  Key here is the enhanced \deletetext{escape}\newtext{first-passage} through a fixed $x_f$ with increasing $\beta$.  
In summary, Fig.~\ref{fig6} demonstrates that, rather than classical OU dynamics or slingshot effects, memory-accumulated positive fluctuations towards the target is the most likely mechanism underlying the escape process through $x_f$.  The transition between classical OU \deletetext{escape}\newtext{first-passage dynamics} and that due to memory-accumulated positive fluctuations is delineated by $\beta = 0$.

\begin{figure}[h]
    \centering
    \includegraphics[scale=0.75]{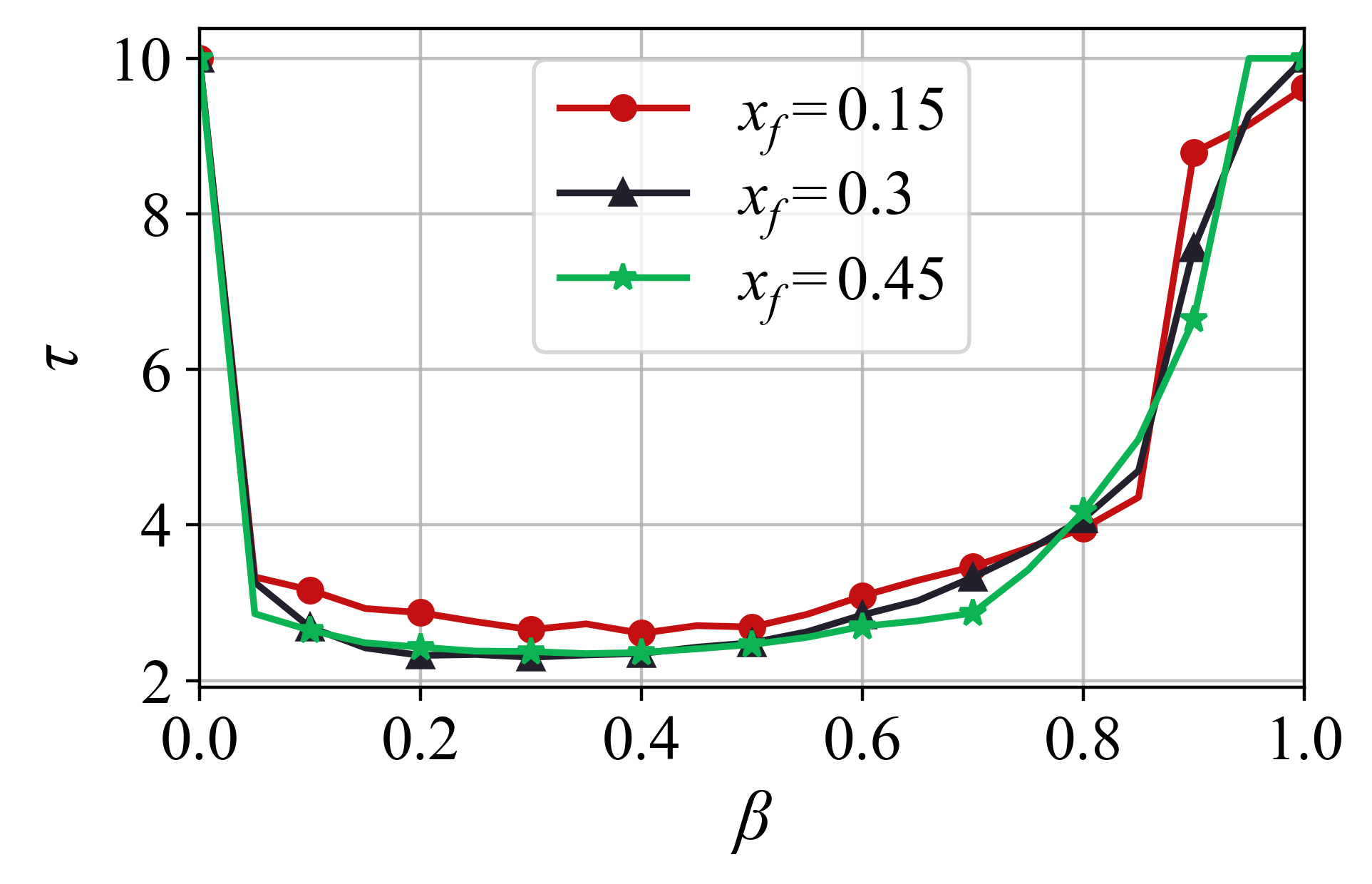}
    \caption{Optimal \deletetext{escape}\newtext{first-passage} time $\tau$ as a function of the feedback parameter $\beta$ for several values of \deletetext{escape}\newtext{first-passage} point $x_f$. For $\beta \leq 0$, $\tau$ diverges and saturates at the numerical cutoff $T=10$. For $\beta > 0$, the optimal time becomes finite, confirming the acceleration of \deletetext{escape}\newtext{first-passage} dynamics. As $\beta \to 1$, $\tau$ increases again and eventually reaches the cutoff value.}
    \label{fig7}
\end{figure}

Finally, in Fig.~\ref{fig7} we plot the optimal \deletetext{escape}\newtext{first-passage} time $\tau$ as a function of $\beta$ for three points $x_f$. This provides further evidence that the global minimum of the action $S_T$ shifts from infinite time for $\beta \leq 0$ (with the proviso of the cut off at $T=10$) to a finite time for $\beta \in (0,1)$. Near $\beta = 0$ the transition appears abrupt, however, further investigation is needed to fully characterize the nature of this transition. The emergence of a finite optimal \deletetext{escape}\newtext{first-passage} time for $\beta > 0$ reflects the role that feedback plays in accelerating the \deletetext{escape}\newtext{first-passage} dynamics through memory-accumulated fluctuations.


Furthermore, for all $x_f$, the optimal \deletetext{escape}\newtext{first-passage} time initially decreases with $\beta$, reaches a minimum, and then grows again for larger $\beta$, eventually reaching the numerical cutoff $T=10$. This suggests that, although feedback initially accelerates \deletetext{escape}\newtext{first-passage dynamics}, the global minimum of the action gradually shifts to longer times as $\beta$ increases. Recall from Section~\ref{subsec:Deterministic} that $\beta=1$ denotes a bifurcation in the deterministic dynamics at which the origin $z_0 = (0,0)$ becomes an unstable fixed point, and small fluctuations can rapidly drive the process away from it, in either direction along the $x$-axis.  Thus, it is difficult to predict the behavior of the optimal time as $\beta \to 1$, but the saturation at $T=10$ in Fig.~\ref{fig7} suggests a divergence in the optimal time, possibly due to the contribution of trajectories escaping to the left.


\section{Conclusion}
\label{sec:Conclusion}

We have investigated the \deletetext{escape}\newtext{first-passage} dynamics of an exemplary non-Markovian stochastic process.  The non-Markovianity is treated by modifying a one-dimensional Ornstein--Uhlenbeck process using a time-averaged feedback wherein the drift 
is influenced by the empirical mean of its own trajectory. Importantly, this model belongs to the broader class of self-interacting diffusions and provides a simple, yet semi-analytically tractable, framework to study the effect of memory feedback on rare events. Beyond its theoretical appeal, such dynamics are relevant in many contexts, including feedback control systems~\cite{Bechhoefer2021} and reinforcement learning algorithms~\cite{Sutton1998}, where history-dependent feedback plays a central role in optimizing task execution and controlling stochastic agents.


We used a weak-noise large deviation approach to derive the equations governing the most probable paths \newtext{that reach a specified position at a given time}, or instantons, and characterized the associated action and quasi-potential. We found that the feedback parameter $\beta$ delineates three qualitatively distinct \deletetext{escape}\newtext{first-passage} mechanisms. When $\beta > 0$, \deletetext{escape}\newtext{first-passage} is driven by the gradual accumulation of positive fluctuations that shift the effective potential and lead to accelerated dynamics through a reduction of the effective energy barrier \newtext{at leading order in the noise}. The optimal \deletetext{escape}\newtext{first-passage} time, which is the time that minimizes the action, also transitions from being infinite for $\beta \leq 0$ to finite for $\beta \in (0,1)$, providing further evidence of accelerated \deletetext{escape}\newtext{first-passage} dynamics. When $\beta < 0$, \deletetext{escape}\newtext{first-passage} occurs either through a slingshot-like mechanism or through an almost-ballistic trajectory, both of which are suboptimal compared to classical OU behavior. Our results establish a clear link between feedback strength and \deletetext{escape}\newtext{first-passage} likelihood, showing that memory can be harnessed to fundamentally reshape the statistics of rare events.


In addition to its basic theoretical relevance, our study suggests more broadly that empirical occupation measures could act as versatile memory devices in feedback-controlled systems. In particular, we suggest that one could encode information on transient rare events into a feedback mechanism that influences future dynamics. This feedback, naturally expressible in terms of the empirical occupation measure, is especially responsive on short time scales, where it adapts quickly to new information. On longer time scales, it relaxes towards the invariant distribution, becoming progressively harder to manipulate. This feature makes it particularly suitable for capturing finite-time rare events and using that memory to modulate a system response in a closed-loop control setting, effectively guiding the process to behave in a desired manner.


Our work suggests a number of immediate questions for future research. \newtext{The model studied here is a continuous-space, continuous-time self-interacting process. It is instructive, however, to draw a parallel with the Elephant Random Walk (ERW) and the Gaussian ERW (GERW)~\cite{Schutz2004,Jack2020}, defined in discrete space--time and in continuous space with discrete time respectively. Despite these differences in construction, their fluctuation paths---featuring giant leaps and long excursions---are reminiscent of the first-passage dynamics we observe for $\beta>0$, where beneficial fluctuations are retained through memory feedback. The analogy, however, cannot be pushed too far: the ERW and GERW couple to the empirical current, whereas our model couples to the empirical density; in particular its mean. This raises the intriguing possibility of a broader ``universality class'' of fluctuation mechanisms in self-interacting processes, extending beyond the specific interaction considered here.}

In designing \newtext{feedback-controlled} systems \newtext{that use self-interaction as a versatile memory device}, both theoretical questions and practical implementations are of interest. A compelling direction is to examine the role of finite-time window memory as a feedback mechanism. Rather than considering a full-time average of the form $t^{-1} \int_0^t K(X_s, X_t) \, ds$, we suggest introducing a feedback based on a time-dependent finite-time window, such as $\lambda(t)^{-1} \int_{t - \lambda(t)}^t K(X_s, X_t) \, ds$. This formulation is truly non-Markovian and, in contrast to the model considered here, does not appear to admit a mapping to a finite-dimensional Markov system. Therefore, within the framework developed in this paper, one could also explore alternative forms of feedback, including for example those governed by the empirical variance or standard deviation of the process, rather than the empirical mean. Finally, it is of interest to analyze the effect of a changing environment by incorporating multiplicative noise, explicitly time-dependent noise or a non-autonomous deterministic backbone. 


\section{Acknowledgments}

FC gratefully acknowledges stimulating discussions with Cristobal Arratia, Ralf Eichhorn, and Supriya Krishnamurthy at Nordita and with Baruch Meerson at Les Houches during the School `Theory of Large Deviations and Applications' held in July 2024. FC is also thankful to Tobias Grafke for insightful discussions and for pointing out the correct final conditions to be used when solving Hamilton's equations in~\eqref{eq:HamiltonEqBoundarySpec}. \newtext{The authors are grateful to Martin Rosinberg for stimulating discussions and for pointing out, after the arXiv submission, a numerical error in the code used to compute the action. This has since been corrected without qualitatively affecting the main results. The authors are also grateful to an anonymous referee for pointing out how to obtain Eq.\ (62).} RD was supported by a scholarship from ENS Paris-Saclay (FR) during his Master 1 internship at Nordita. We acknowledge partial support from the Swedish Research Council under Grant No.\ 638-2013-9243 and from EPSRC under Grant No.\ EP/V031201/1.

\section*{References}

\bibliographystyle{ieeetr}
\bibliography{mybib}

\newtext{
\section{Appendices}

\subsection{Derivation of boundary value problem for mean first-passage time in Eq.\ \eqref{eq:BackwardMeanEscapeTime}}
\label{ref:appa}

The boundary value problem in \eqref{eq:BackwardMeanEscapeTime} can be obtained by adapting the derivation of result 7.1 in~\cite{Pavliotis2014} by considering a time-dependent generator, such as that in \eqref{eq:InfinitesimalGenMarkovFeedback}. Given the definition of mean first-passage time in \eqref{eq:MeanFPTFeedback} and in terms of the survival probability \eqref{eq:FirstPassageTimeDistr}, we can write
\begin{align}
    \bar{T}_s(z_s) &= \mathbb{E}_{s,z_s} \left[ T_s(z_s) \right] \\
    &= \int_s^{+ \infty} f_{s,t}(z_s) (t-s) \, dt \\
    &= - \int_s^{+ \infty} \frac{\partial Q_{s,t}(z_s)}{\partial t} (t - s) \, dt \, ,
\end{align}
which, by first integrating by parts and by then using \eqref{eq:Survival}, can be re-written as
\begin{align}
    \bar{T}_s(z_s) &= \int_s^{+ \infty} Q_{s,t}(z_s) \, dt \\
    &= \int_s^{+\infty} \left( \int_{-\infty}^{x_f} \int_{-\infty}^\infty \rho(z,t|z_s,s) \, d\bar{x} \, dx \right) \, dt \, .
\end{align}
Then, by introducing the semigroup associated to the time-dependent infinitesimal generator \eqref{eq:InfinitesimalGenMarkovFeedback} and using the initial condition in~\eqref{eq:ForwardTransitionProba} we find
\begin{align}
    \bar{T}_s(z_s) &= \int_s^{+\infty} \left( \int_{-\infty}^{x_f} \int_{-\infty}^\infty e^{L_s^\dagger (t-s)} \delta(z_s-z) \, d\bar{x} \, dx \right) \, dt \\
    &= \int_s^{+\infty} \left( \int_{-\infty}^{x_f} \int_{-\infty}^\infty \delta(z_s-z) \left( e^{L_s (t-s)} 1 \right)(z) \, d\bar{x} \, dx \right) \, dt \\
    &= \int_s^{+\infty} \left(  e^{L_s (t-s)} 1 \right) (z_s) \, dt \, .
\end{align}

Applying $L_s$ on both sides, we obtain
\begin{align}
    L_s \bar{T}_s(z_s) = \int_s^{+\infty} \frac{d}{d t} \left( e^{L_s(t-s)} 1 \right)(z_s) \, dt \, ,
\end{align}
which, after integrating by parts, yields
\begin{equation}
    L_s \bar{T}_s(z_s) = -1 \, , \qquad \text{for } z_s \in (-\infty, xf) \times (-\infty, x_f) \, ,
\end{equation}
and $\bar{T}_s(z_s) = 0$, for $z_s \in x_f \times (-\infty,\infty)$.
}

\end{document}